\documentclass[a4paper,12pt]{article}
 \usepackage{graphicx}
 \usepackage{latexsym}
 
\newfont{\ftsect}{cmbx12 scaled\magstep2}
\newfont{\fttitle}{cmbx12 scaled\magstep2}
\newfont{\ftauthor}{cmbx12 scaled\magstep1}
\newfont{\ftabstract}{cmsl12}
\newfont{\texte}{cmr12 scaled\magstep1}
\newfont{\xbf}{cmbx12 scaled\magstep1}
\newfont{\xit}{cmti12 scaled\magstep1}
\newfont{\xsl}{cmsl12 scaled\magstep1}
\newfont{\xtt}{cmti12 scaled\magstep1}
\newfont{\xsmall}{cmr12}
\newfont{\smallit}{cmti12}

\renewcommand{\sl}{\xsl}

\renewcommand{\small}{\xsmall}
         
\newcommand{\ds}{\displaystyle}
\newcommand{\rr}{{\bf R}}
\newcommand{\ri}{{\rm i}}
\newtheorem{defin}{Definition}
 
\makeatletter
\@addtoreset{equation}{section}
\makeatother
\renewcommand{\subsection}[1]{\vspace{1mm}
 
\addtocounter{subsection}{1}\noindent
{\bf \thesubsection \ \ #1}
\vspace{1mm}
 
\noindent}

\renewcommand{\title}[1]{\null\vspace{28mm}

\begin{center}{\fttitle{\bf #1}}\end{center} }
\renewcommand{\author}[1]{\vspace{1mm}
 
\begin{center}{\ftauthor #1}\end{center} }
\newcommand{\address}[1]{\vspace{-5mm}
 
\begin{center}{\small  #1}\end{center} }
\renewcommand{\abstract}[1]{\vspace{15mm}

\noindent{{\smallit Abstract.} {\small #1}}
\large\texte  
}
\def\ftoday{{\sl  \number\day \space\ifcase\month
\or Janvier\or F\'evrier\or Mars\or avril\or Mai
\or Juin\or Juillet\or Ao\^ut\or Septembre\or Octobre
\or Novembre \or D\'ecembre\fi
\space  \number\year}}

\newcommand{\sla}{\raise.15ex\hbox{$/$}\kern -.57em}
\newcommand{\Sla}{\raise.15ex\hbox{$/$}\kern -.70em}

\newcommand{\complex}{{\kern .1em {\raise .47ex
\hbox {$\scriptscriptstyle |$}}
    \kern -.4em {\rm C}}}
\newcommand{\real}{{{\rm I} \kern -.19em {\rm R}}}
\newcommand{\rational}{{\kern .1em {\raise .47ex
\hbox{$\scripscriptstyle |$}}
    \kern -.35em {\rm Q}}}
\renewcommand{\natural}{{\vrule height 1.6ex width
.05em depth 0ex \kern -.35em {\rm N}}}

\newcommand{\pa}{\partial}

\newcommand{\twiddle}{\lower.9ex\rlap{$\kern -.1em\scriptstyle\sim$}}

\newcommand{\equ}[1]{(\ref{#1})}
\newcommand{\eq}{\begin{equation}}
\newcommand{\eqn}[1]{\label{#1}\end{equation}}
\newcommand{\eea}{\end{eqnarray}}
\newcommand{\eqa}{\begin{eqnarray}}
\newcommand{\eqan}[1]{\label{#1}\end{eqnarray}}
\newcommand{\ba}{\begin{array}}
\newcommand{\ea}{\end{array}}
\newcommand{\eqac}{\begin{equation}\begin{array}{rcl}}
\newcommand{\eqacn}[1]{\end{array}\label{#1}\end{equation}}

\newcommand{\remarks}{\bigskip 
 
   \noindent{\bf Remarks:} \begin{enumerate}}
\newcommand{\skramer}{\end{enumerate}}

\begin{document}
 
\hfill LYCEN 9754 

\hfill December 1997

\title{ABOUT \\[3mm]
SYMMETRIES IN PHYSICS}
\begin{center}
{Dedicated to H. Reeh and R. Stora}\footnote{\small I 
wish to dedicate these notes to my diploma and Ph.D.
supervisors H. Reeh and R. Stora who devoted a major part
of their scientific work to the understanding, description
and exploration of symmetries in physics.}
\end{center}
\author{Fran\c cois Gieres}
\address{Institut de Physique Nucl\'eaire de Lyon,
 IN2P3/CNRS,  
         Universit\'e Claude Bernard \\
         43, boulevard du 11 novembre 1918, 
         F - 69622 - Villeurbanne CEDEX}	

\abstract{
The goal of this introduction to symmetries is
 to present some general ideas, 
 to outline the fundamental concepts and results
of the subject 
and to situate a bit the following lectures of this school. 
[These notes represent the write-up 
of a lecture presented at the fifth {\em S\'eminaire Rhodanien de
Physique} ``Sur les Sym\'etries en Physique" held at Dolomieu 
(France), 17-21 March 1997.  Up to the appendix and the graphics,  
it is to be published in {\em Symmetries in Physics}, F. Gieres, M.
Kibler, C. Lucchesi and O. Piguet, eds. (Editions Fronti\`eres,
1998).]}  

\thispagestyle{empty}

\newpage

\thispagestyle{empty}

\begin{centerline}
{\bf {\huge Contents}}
\end{centerline}

\vspace{1cm}

\noindent
{\bf 1 $\; $ Introduction}  \dotfill 1

\bigskip 
\noindent
{\bf 2 $\; $ Symmetries of geometric objects} \dotfill 2 
 
\bigskip 
\noindent
{\bf 3 $\,$ Symmetries of the laws of nature} \dotfill 5

{\bf 1 $\, $ Geometric (space-time) symmetries} \dotfill 6

{\bf 2 $\,$ Internal symmetries} \dotfill 10

{\bf 3 $\, $ From global to local symmetries} \dotfill 11

{\bf 4 $\, $ Combining geometric and internal symmetries} \dotfill 14

{\bf 5 $\, $ Duality symmetries} \dotfill 15

{\bf 6 $\, $ Miscellaneous} \dotfill 16

\bigskip 
\noindent
{\bf 4 $\; $ The mathematical description of symmetries and their
implementation  in physical theories} \dotfill 16

{\bf 1 $\, $ Mathematical description : (Lie) groups and algebras}
\dotfill 16

{\bf 2 $\, $ Physical implementation : representations} \dotfill 18

{\bf 3 $\, $ Generalizations} \dotfill 18

\bigskip 
\noindent
{\bf 5 $\; $ Implications of symmetries for the formalism, the
results and the structure of physical theories} \dotfill 20

\bigskip 
\noindent
{\bf 6 $\; $ Different manifestations of symmetries} \dotfill 22

{\bf 1 $\; $ Broken symmetries} \dotfill 23

{\bf 2 $\; $ Miscellaneous and some important asymmetries} \dotfill
26

\bigskip 
\noindent
{\bf 7 $\; $ Conclusion} \dotfill 27

\bigskip 

{\bf A.1 $\, $ On (Lie) groups and algebras} \dotfill 28

{\bf A.2 $\, $ About representations} \dotfill 32


\newpage

\setcounter{page}{1}

   \section{Introduction}
 
`Symmetric' objects are aesthetically appealing 
and fascinating for the human mind. But what 
does symmetric mean?
The original sense of the Greek word {\em symmetros} is 
`well-proportioned' or `harmonious'.
In his classic work on symmetry \cite{weyl}, H. Weyl puts it
the following way :   
``Symmetry denotes that sort of concordance of several 
parts by which they integrate into a whole. {\em Beauty} is
bound up with symmetry." 
This fact might account for the  
omnipresence of symmetries 
in nature and our description of it, if we adhere to the views  
of d'Arcy Thompson 
\cite{darcy} :
``The perfection of mathematical beauty is such that 
whatever is most beautiful and regular is also found 
to be most useful and excellent." 

In less poetic terms (and with a thought for the great 
cathedral builders of the Renaissance), 
we could say that 
stable complex systems 
are best created by assembling in a regular way symmetric
constituents blocks. Thus, it is natural that  
symmetries manifest themselves at all levels     
- microscopic and macroscopic - 
in our world as we see it, as we comprehend it and 
as we shape it. 

\smallskip 

\noindent
{\bf Outline of the notes}

These notes are intended to be of an elementary level,  
though some remarks refer to some more specific knowledge of 
physics or mathematics. 
(For the more technical parts of the notes, I assume   
some familiarity  with 
the algebraic tools 
 of the subject which the reader should remember from 
quantum mechanics :
these are the notions  of group, Lie group, Lie algebra and the
representations  of these algebraic structures. 
These concepts are briefly reviewed in section 4 
and for the reader's convenience, we have summarized the basic definitions
together with some illustrative
examples in an appendix.) In my write-up, 
I have preferred to maintain the informal 
character of the lectures rather than making 
attempts for formal rigor.   

In the next two sections, we successively define
and  classify 
 symmetries 
of geometric objects in space (which may be referred to as
{\em spatial symmetries}) 
and symmetries of the laws of nature ({\em physical symmetries}). 
In section 4, we recall how symmetries are described 
in mathematical terms (groups, algebras,...)
and how these mathematical structures 
are implemented in physical theories (representations,...). 
Thereafter, it is shown  how the presence of  
symmetries affects physics at different levels :
the formalism, the predictions and the general structure of 
theories. 
In conclusion, different manifestations
 of symmetries
are indicated (approximate symmetries, 
broken symmetries,...).

\newpage

\noindent 
{\bf About the literature}

 Though I tried to convey a flavor of the subject, 
it was neither possible nor intended 
to present an exhaustive treatment of it.
For further information, 
the interested reader is referred
to the many excellent and fascinating textbooks which are 
largely or completely devoted to symmetries : 
 a selection of them is  
given in the bibliography 
\cite{feynman}-\cite{penrose}. 
We particularly mention 
the wonderful introductions to symmetries  
given by H. Weyl \cite{weyl} 
and R. Feynman \cite{feynman} in series of lectures 
addressed towards a general audience.   
In preparing these notes, I repeatedly used the books 
of Tarassow \cite{tara} and Genz/Decker \cite{genz}
which I recommend 
wholeheartedly together with the pleasant and somewhat 
encyclopedic monograph of Sivardi\`ere \cite{siva} (though 
I am afraid none of these 
texts is available 
in English ...).


\section{Symmetries of geometric objects}

\begin{figure}[h!]
\centerline{\includegraphics*[height=5cm,angle=0]{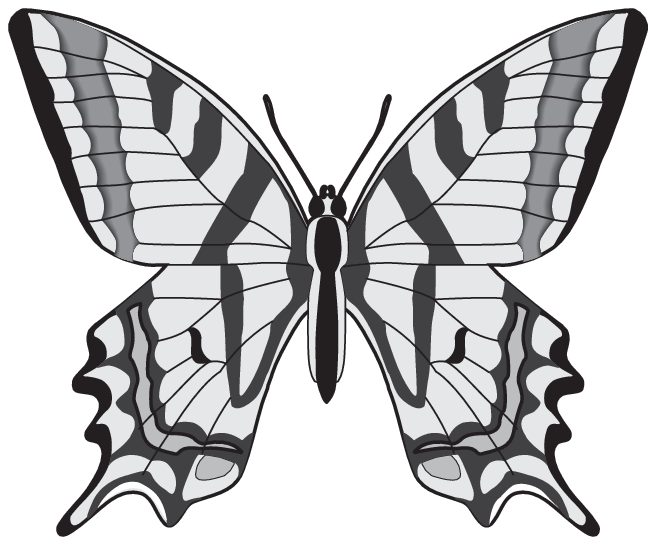}}
\end{figure}

\bigskip

In everyday life, symmetry usually means 
{\underline{left-right (mirror or bilateral)
  sym-} \underline{metry}.
For an object in a plane, this means that there exists 
a {\em symmetry line} and for an object in space, it amounts 
to the existence of a {\em symmetry plane}: if you exchange the 
two sides of the object, it looks exactly the same. 
This symmetry is realized - at least approximatively - 
for human beings and higher animals. (Henceforth, it is also 
the first symmetry of which we became
 aware in our personal life.)

One may wonder whether there is any reason for the occurrence of 
this symmetry for humans and animals. Obviously, there are 
two preferred directions for all of these beings : 
on one hand, there is the direction of motion (which is used
when looking for food, attacking the enemy or joining friends)
and on the other hand, there is the direction of 
gravity to which everybody is subject to. Altogether 
these two directions define a plane of symmetry in space. 

What can we learn from these simple reflections?
They show that symmetries usually reflect some intrinsic 
properties 
or characteristics of {\em objects} and of the {\em space}
 to which they belong.
And they indicate that symmetries in nature are often realized 
as {\em nearby} ({\em approximate}) symmetries rather than
in an {\em exact} way.
It should be noted that, in general,  
\underline{symmetries in the living nature}
only manifest themselves if they are privileged from    
the point of view of selection \cite{eigen}.

The interpretation of our initial example is confirmed 
by looking at the plant world : for plants and trees, there 
is only one distinguished direction, namely the one given 
by gravity, all horizontal directions being equally well 
suited for absorbing oxygen, light or humidity. 
Consequently, plants and trees admit one symmetry axis
and thus have \underline{rotational symmetry}. 
Trees with many branches approximately have a {\em continuous} 
rotational symmetry around the vertical axis of their trunk, 
while flowers rather have a {\em discrete} rotational symmetry.
In fact, many flowers have five petals and therefore admit a 
discrete rotational symmetry of fifth order : they are transformed 
into themselves when turned by an angle 
$n \, {360^{\circ} \over 5} \ ({\rm with} \; n \in {\bf Z})$ around their 
symmetry axis. 

Again one may wonder whether there is a deeper reason for this  
symmetry of order five that many flowers share?
Simple models for the growth of flowers 
show that the number of petals is a {\em Fibonacci number} 
($1,2,3,5,8,13,..$) \cite{conway}. (Such models  	
 also explain the  manifestation of  these
numbers in  the symmetric arrangement of  
sunflowers seats 
or of branches around the stem of a tree,...)  
Should you notice many flowers with  
six petals on your next outdoor trip, then you should not 
conclude that the models we mentioned are wrong :  
the growth conditions for the buds of  
these flowers are different and 
such that their petals  get organized 
as two generations 
of $3$ petals each, $3$ being a Fibonacci number 
too!

A very different and somewhat metaphysical argument
 in favor of the number $5$ is referred to in  
\cite{tara}. The five petals of a flower span a 
regular pentagon and if you try to put as many as possible
of these together (i.e. pave or tile a plane with them), then 
some spaces will 
remain uncovered
 between these pentagons - see figure below.
  In fact, such a tiling 
 can be achieved with regular triangles, 
squares or hexagons, but {\em not} with pentagons. 
One can argue that by choosing a fifth-order symmetry,
 the flowers 
 try to fight for survival and protect themselves
against crystallization, the first step of which would consist 
of getting `trapped' in a crystal lattice.     
Though this remark should not be taken too seriously, 
it reveals an important fact concerning symmetries in nature
 \cite{bronowski} :

\begin{tabular}{|p{15cm}|}
\hline  
Only certain symmetries 
are supported by the space in which we live. 
\\
\hline
\end{tabular} 

\noindent 
This does not only apply to patterns created by men, 
but to all regular structures occurring in nature.  
(Our example of pentagons refers to planar domains,
 but we can also think about our three-dimensional 
space :
there are only {\em five  Platonic bodies}, i.e. regular 
polyeders like the cube or tetraeder.)
Even the mathematical tools we use to describe physical phenomena 
are reflections of the symmetries 
of space. For instance, 
the {\em Pythagorean theorem} would not hold as such
if space were to have 
other symmetries.

\begin{figure}[h!]
\centerline{\includegraphics*[height=6cm,angle=0]{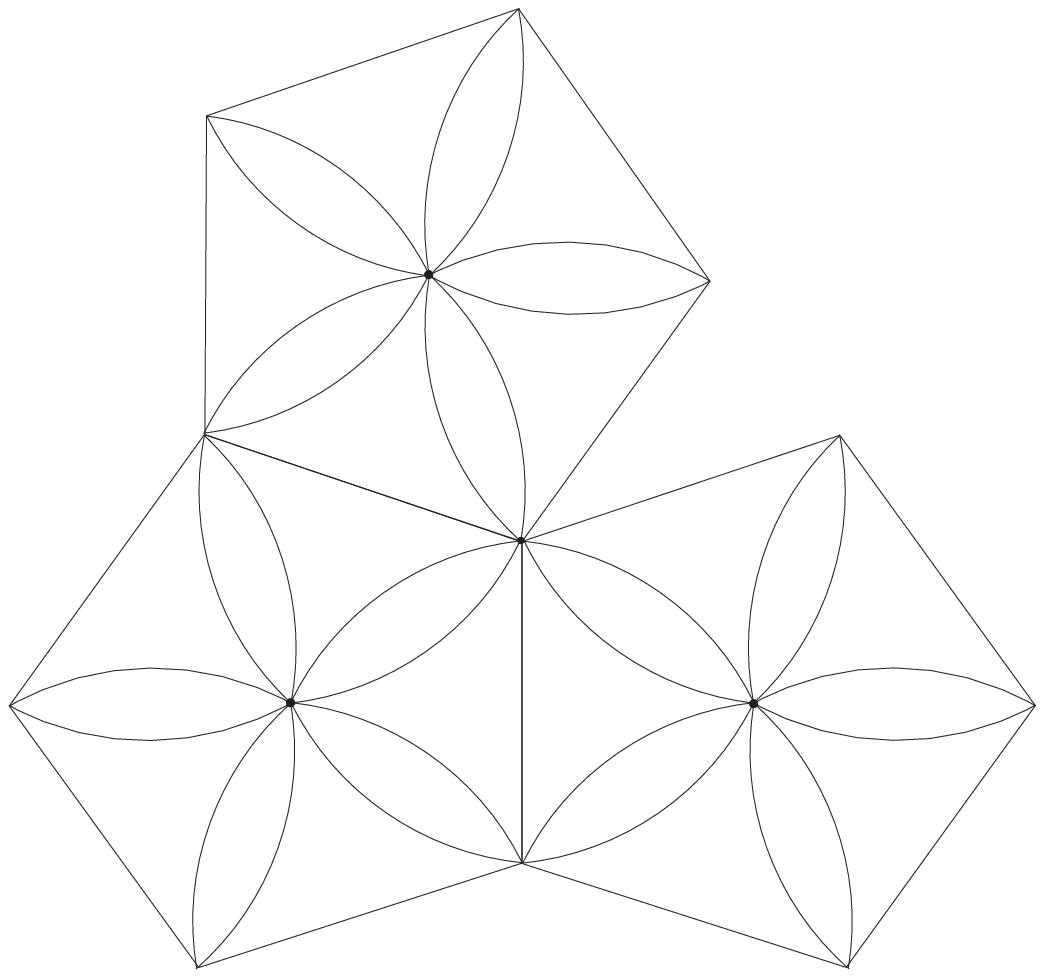}}
\end{figure}

The issue of tilings brings us to the subject of 
 \underline{symmetries in art} \cite{shub},
 e.g. the famous tilings of the 
Alhambra,
ornamental symmetry,  
Escher's drawings \cite{escher},
Penrose tilings \cite{siva},... 
The construction of these regular patterns is based on 
\underline{translational
symmetry} and combinations of all 
spatial symmetries introduced so far.

The paradigm, or perfect example, 
of symmetries is definitely given by the 
\underline{crystal}. 
For a discussion of regular and \underline{quasicrystals},
 we refer 
to \cite{siva,crystal} and the lectures of 
J.P. Gazeau in this volume.

After contemplating all these  examples,
the reader may still wonder :

\smallskip 
\noindent 
{\bf What {\em is} a symmetry?}

Roughly speaking, 
an object 
is symmetric (has a symmetry) if there is something you can 
do to it,
so that after you finished doing it, it looks exactly 
the same way it did before. In other words, a      
{\em symmetry transformation} of a geome\-tric object 
(which is part of a plane or of space) is a 
transformation of the object 
whose realization (effect) is impossible to detect. 
For instance, for 
the following geometric figures, 
 it is not possible 
to notice the effect of  a rotation
 by an angle of, respectively,  $180^{\circ}$, 
$90^{\circ}$  or  an  arbitrary value 
around their symmetry axis.  

\begin{picture}(300,90)(0,0) 
\put(20,30){\framebox(100,50)}
\put(210,30){\framebox(50,50)} 
\put(370,55){\circle{70}}
\put(70,55){\circle*{2}}
\put(235,55){\circle*{2}} 
\put(370,55){\circle*{2}}
\end{picture}

\vspace{-5mm}

\noindent 
The more symmetries an object admits, the
{\em more symmetric} it is.

In order to formulate these ideas more precisely 
from the mathematical point of view \cite{knoerrer, penrose}, 
we recall that an 
\underline{isometry} 
of a geometric object in Euclidean space 
is a transformation of it   
which {\em preserves the distances}. Examples are 
given by rotations, translations or  reflections. 
\begin{defin}
A \underline{symmetry} or \underline{symmetry transformation}
 of a geometric object in Euclidean space 
is an isometry which maps the object onto itself. 
If an object admits a 
certain symmetry, 
it is said to have this \underline{invariance}.
\end{defin}
The set of all symmetry  transformations of an object
represents a group (the
group multiplication being the consecutive application of 
transformations) : this is the   
\underline{symmetry group of the object}.

\section{Symmetries of the laws of nature} 

We can fit the definition given for objects in space 
to the present case: loosely speaking, a 
{\em symmetry of a law of nature} means that
 there is something
we can do to a  physical law - or rather to our way
 of representing it - which makes no difference and 
leaves everything unchanged in its effects.    
To be more precise, we
 consider a physical system described by a {\em law} 
or, more specifically, by  
some {\em equations} involving a certain number of 
{\em variables} which may (or may not) represent
 directly observable quantities, 
and which possibly depend on the {\em space-time coordinates}.
E.g. we consider the propagation of an impulse with
 the speed of light $c$,   
as described by the     equation 
$\Box \varphi (\vec r ,t) =0$ where 
$\Box  \equiv {1 \over c^2} \, \pa_t ^2 - (\pa_x^2 +\pa_y^2
+ \pa_z^2)$ 
denotes the wave operator. 

\begin{defin}
A \underline{symmetry transformation} of a physical law
is a change of the variables and/or space-time 
coordinates
(in terms of which it is formulated)
such that the equations describing the law 
have the same form in terms of the new 
variables and coordinates as they had in terms
  of the old ones. 
One says that the equations 
preserve their form or that they are \underline{covariant 
with respect to the sym-} 
\underline{metry transformation}. 
\end{defin}

Thus, the realization of a symmetry transformation is 
impossible to detect.
In our example of the wave equation, we may apply a 
\underline{Lorentz transformation}
 (a boost) with velocity $\vec v = (v,0,0)$ : under this 
operation, the space-time coordinates $(\vec r,t)$ become 
\eq
x^{\prime} = \gamma ( x - vt) \quad , \quad      
y^{\prime} = y \quad , \quad 
z^{\prime} = z \quad, \quad 
t^{\prime} = \gamma ( t - {v \over c^2} x) 
\ \ , 
\eqn{lor}
where $\gamma = (1 -v^2 / c^2 )^{-1/2} $, 
but the wave operator       
is invariant under these  transformations, 
i.e. $\Box^{\prime} = \Box$. This result,  
first found by H. Lorentz, and the fact that 
the scalar function $\varphi$ has the same numerical value
in the transformed and original reference frames, i.e. 
$\varphi^{\prime} (\vec r ^{\, \prime} , t^{\prime} ) 
= \varphi (\vec r ,t)$, 
imply that the wave equation has exactly the same form 
in both reference frames.
In physical terms: the wave propagates in the same way 
and with the same velocity $c$ in two inertial frames that 
are in uniform motion relative to each other.
With another phrasing we can say that,  
 by analyzing the wave propagation 
in a spaceship, an observer cannot tell whether 
this spaceship is at rest or in uniform motion relative to the 
stars (unless he looks outside of its windows).

In the next sections, we  
will be concerned with physics in {\em flat, four-dimen-}
{\em sional} 
space-time. Curved space
and symmetries in  
other-dimensional spaces,  which are relevant for 
statistical mechanics or string theory, 
will only be  commented upon. 

For the {\em classification of  symmetries}, one distinguishes 
between those which operate on space-time coordinates,  
the so-called \underline{geometric symmetries} and those
which do not affect them, the \underline{internal symmetries}. 

\medskip

\subsection{Geometric (space-time) symmetries}
\vspace{-9mm}

The basic operations are the following ones \cite{roemer} :
\eq
\begin{array}{ll}
\mbox{Translation in time :} &
\, t^{\prime} = t + a^0 \\
\mbox{Translation in space :} & 
\vec{r} ^{\, \prime} = \vec r + \vec a \\
\mbox{Rotation in space :} &
\vec r ^{\, \prime} = R \, \vec r 
\ \ ,   
\end{array} 
\eqn{geo}
where $a^0 \in \rr, \ \vec a \in {\bf R}^3$, $R$ being
 a $3\times 3$ 
rotation matrix.
These transformations preserve the form of 
time evolution equations, e.g. of the equation 
$m \ddot{\vec r} = \vec 0$ in classical
mechanics.    
(Thus, some potential symmetries for 
objects in space are   basic
symmetries of the laws of nature.)
The origin of these fundamental invariances  can be traced back
 to {\em intrinsic properties of space and
 time} - see the table below. 

Since the changes of coordinates  \equ{geo}
are {\em continuous}
symmetry transformations in the sense that they are parametrized 
in a smooth way by real numbers, we can apply the    
famous \underline{theorem of E. Noether} :

\begin{tabular}{|p{16cm}|}
\hline  
\smallskip 
Covariance of the equations of motion with respect to  a 
continuous transformation with $n$ parameters 
implies the existence 
of $n$ conserved quantities (`conserved charges' or `integrals
of motion'), i.e. it implies  
\underline{conservation laws}.  
\smallskip 
\\
\hline
\end{tabular} 

\noindent 
We have the following 
correspondence in the present case :   

\begin{tabular}{|l|l|l|}
\hline 
Property  & Invariance of equations & Conserved quantity \\ \hline \hline 
homogeneity of time & time translation invariance & energy \\ \hline  
homogeneity of space & translational invariance & momentum \\ \hline  
isotropy of space & rotational  invariance & angular momentum \\ \hline  
\end{tabular}

In {\em non-relativistic} theories like Newtonian mechanics 
or usual quantum mechanics,
the time evolution equations are also covariant 
with respect to the  
\underline{Galilean transformation}
\begin{equation}
\vec r ^{\, \prime} = \vec r - \vec v \, t  
\qquad {\rm where} \ \; \vec v \in {\bf R}^3 
\ \ .
\end{equation}
In {\em relativistic} theories like electromagnetism,  
this symmetry transformation is to be replaced by the 
Lorentz transformation \equ{lor}. 

In order to discuss the relativistic symmetries 
from a formal point of view,  
it is convenient to introduce 
{\em Minkowski space} $(\rr^4, (ds)^2)$, i.e. 
the real vector space $\rr^4$ parametrized by space-time 
coordinates $x = (x^{\mu}) _{ {\mu} = 0,1,2,3 } 
= (ct , \vec r \, )$ and equipped with the metric 
\eq 
(ds)^2 = (dx)^2 \equiv c^2 dt^2 - (d\vec r \, )^2 
\ \ . 
\eqn{mink}
By definition, the \underline{Lorentz group} $L$ 
is the set of all real $4\times4$ matrices $\Lambda$ 
which leave the 
Minkowski metric \equ{mink} 
invariant, i.e. all linear coordinate transformations 
$x \to x^{\prime} = \Lambda x$ such that $(dx^{\prime}) ^2
= (dx)^2$. These include the rotations and Lorentz 
boosts, but also the operations of \underline{parity}
($\vec r^{\, \prime} = - \vec r \,$) and \underline{time 
reversal} ($t^{\prime} = -t$) \cite{sachs}.
The latter transformations are referred to as {\em discrete} 
- as opposed to {\em continuous} - symmetries.

The space-time translations $x \to x +a$ also leave 
the Minkowski metric invariant, because  
$d(x+a) = dx$. They form a group,  
the \underline{group of translations} ${\cal T}_4$, i.e.  
the set of all mappings $T_a$ (with  
$a = (a^0, \vec a \, ) \in {\bf R}^4$ ) defined by 
\begin{eqnarray}
T_{a} \ : \ & {\bf R}^4 &  \longrightarrow \ {\bf R}^4 
\nonumber 
\\
 & x &  \longmapsto \ T_{a} ( x  ) = x + a
\ \ , 
\nonumber
\end{eqnarray} 
the group multiplication being the 
composition :
$T_{a} \circ T_{b} = T_{a +  b}$. 
By combining Lorentz transformations and translations, we obtain 
the \underline{Poincar\'e
group}
\eq
P = L \subset \! \! \! \! \! \! \times {\cal T}_4 
\ \ , 
\eqn{poi}
whose elements $(\Lambda , T_a)$ act on $x \in \rr^4$ 
according to $x \to  \Lambda x +a$. 
They represent {\em isometries of Minkowski space}. 

An \underline{elementary particle} is an object whose nature  
does not change when it is translated in space or time, 
when it is rotated or seen from an observer in uniform 
motion relative to it. These considerations led E. Wigner to postulate 
that the quantum mechanical  
states of such a particle should belong to a Hilbert 
space carrying a certain representation of the Poincar\'e
group (cf. appendix A.2).
 Thus, the very definition of   
an  elementary particle is based 
on the geometric symmetries and so is the whole 
classification of particles according to their mass 
and spin \cite{weinberg}. A particle may be viewed 
as the `quantum' of a \underline{classical relativistic   
 field} $(A_1 (x),...,A_n (x))$, i.e.   
a collection of 
space-time functions $A_i (x)$  having specific transformation
properties with respect to {\em Poincar\'e} transformations. 
The way in which these fields can 
{\em interact} with each other, is strongly restricted 
by the requirement of relativistic covariance
of the equations of motion.
In summary : 

\begin{tabular}{|p{16cm}|}
\hline  
The Poincar\'e group is an invariance group of all 
relativistic theories and thereby  determines
their general structure to a large extent. 
\\
\hline
\end{tabular}

It should be noted that 
continuous symmetries in physics are often formulated 
in terms of {\em infinitesimal} rather than {\em finite}
 transformations, 
i.e. one considers the {\em Lie algebra}
rather than the {\em Lie group} of transformations
 - see the appendix for an illustration concerning    
rotations.  

One may wonder whether there are more general groups
(or algebraic structures) containing  
the Poincar\'e group as a subgroup, which are 
of physical interest.
Indeed, there are such
extensions of the Poincar\'e group, two of which will 
now be presented.

\vspace{-2mm}
  
\subsubsection{The conformal group}
\vspace{-3mm}

\underline{Conformal transformations} of 
Minkowski space are
coordinate transformations $x \to x^{\prime} (x)$ 
which are such that the induced change of the metric is 
a rescaling by a positive function :
\eq
(ds^{\prime})^2 = {\rm e}^{\Omega (x)} \; ds^2 
\ \ , 
\eqn{cg}
where $\Omega$ is a smooth, 
real-valued function\footnote{\small More 
generally, one can introduce conformal coordinate 
transformations on 
a (pseudo-) Riemannian manifold, i.e. 
a manifold  equipped with a (pseudo-) Riemannian metric.}.
Geometrically, this means that a conformal transformation 
{\em preserves the angles} in magnitude and direction,
though it may locally   
 change the distances in a smooth way. Of course,
 transformations
with this property are essential for the design 
of geographic maps ; remarkably enough, it was already 
realized by Ptolemy that the {\em stereographic projection} 
of a sphere onto a plane represents a conformal mapping.

The set of all conformal transformations
of Minkowski space is called its
\underline{conformal group}.
Of course, this group contains the Poincar\'e
transformations for which $\Omega = 0$.  
But it also involves \underline{dilatations} ({\em rescalings}
 or  {\em scale transformations}) by 
a positive, constant factor ${\rm e}^{\lambda}$ :
$$
x^{\prime} = {\rm e}^{\lambda} \, x   
\ \ \Longrightarrow \ \  
(ds^{\prime})^2 = {\rm e}^{\lambda} \; ds^2 
\ \ . 
$$
Besides, it  contains the 
so-called {\em special conformal transformations} \cite{dubrovin}.

\underline{Conformal field theories}, i.e. 
field theories admitting the conformal group as    
symmetry group have been intensively studied  
during the last decades. Their relevance 
for particle physics is severely 
limited by the fact that the presence of a massive 
particle in such models 
implies the existence of a {\em continuous mass 
spectrum} \cite{jackiw}.   
However, 
scale transformations play an important r\^ole for the description 
of \underline{critical phenomena} (phase transitions)
 in statistical physics \cite{sewell}. Furthermore, 
two-dimensional conformal field theories \cite{conf} 
 are 
at the very heart of \underline{string theories}
which are currently viewed as the most promising candidates
for a unified quantum theory of all fundamental interactions
of nature \cite{string}.

\vspace{-2mm}

\subsubsection{The super Poincar\'e algebra}
\vspace{-3mm}

Let ${\rm Lie} \, P$ denote the Lie algebra associated 
to the Poincar\'e group, i.e. the set of 
infinitesimal translations, rotations and boosts. 
The \underline{super Poincar\'e algebra}
\eq
{\cal G}_{\overline 0}
\oplus {\cal G}_{\overline 1} \ \equiv \ 
{\rm Lie} \, P \ \oplus \{ {\rm SUSY} \} 
\eqn{susy}
is a \underline{Lie superalgebra} 
(or \underline{${\bf Z}_2$-graded Lie algebra}) 
which roughly means the following 
 \cite{corwin,sohnius}.  
The elements of the vector spaces 
${\cal G}_{\overline 0}$ and    
${\cal G}_{\overline 1}$ are assigned     
a degree or parity $0$ and $1$, respectively :
therefore, they are 
referred to as {\em even} and {\em odd} or as {\em bosonic}
and {\em fermionic} elements,  respectively, 
and one often writes ${\cal B} = {\cal G}_{\overline 0}$
and ${\cal F} = {\cal G} _{\overline 1}$.    
Instead of the usual Lie commutator
$[A, B]\equiv AB-BA$,   one has a `${\bf Z}_2$-graded' 
commutator $[A , B\}$ which has the properties of 
 an anticommutator $\{ A , B\} \equiv AB+BA$ 
if $A$ and $B$ are both fermionic 
and otherwise those of a commutator.
Schematically, one has   
\eq
\fbox{\mbox{$
\ {[ {\cal B}, {\cal B} ]} = {\cal B} 
\quad , \quad 
{[ {\cal B} , {\cal F} ]} = {\cal F} 
\quad , \quad 
\{ {\cal F} , {\cal F} \} = {\cal B} 
\ $}}
\ \ . 
\eqn{schema}
In our example of the super Poincar\'e algebra, the relation 
${[ {\cal B}, {\cal B} ]} = {\cal B} $
summarizes the commutation relations 
of the Poincar\'e algebra ${\cal B} = {\rm Lie} \, P$ and 
${\cal F} = \{ {\rm SUSY} \}$ is spanned   
by the so-called \underline{supersymmetry generators}
$(Q_{\alpha})_{\alpha =1,..,4}$: for these 
 elements, the relation
$\{ {\cal F} , {\cal F} \} = {\cal B}$
 takes the explicit 
form 
\eq
\fbox{\mbox{$
\ \{ Q_{\alpha} , Q_{\beta} \} = \sum_{\mu =0}^{3} 
f_{\alpha \beta}^{\mu} \, P_{\mu}
\ $}}
\eqn{susyal}
where $f_{\alpha \beta}^{\mu} \in {\bf C}$ are 
structure constants and $(P_{\mu})_{\mu =0,..,3}$ the 
generators of space-time translations (i.e. $P_{\mu}
= i \partial _{\mu}$ in a representation by differential 
operators).  
In essence, the  
last relation states that

\begin{tabular}{|p{16cm}|}
\hline  
The supersymmetry generators are `square roots' of 
translation generators.
\\
\hline
\end{tabular}

By introducing 
$4N$ (with $N=2,3,...$ ) supersymmetry 
generators $Q_{\alpha}^i$ (where $i=1,..,N$), one can define the 
\underline{$N$-extended
super Poincar\'e algebra}.

\underline{Supersymmetric quantum mechanics} \cite{junker}
and \underline{supersymmetric field theories} 
\cite{sohnius, haber}
represent realizations of the super {\em Poincar\'e} algebra
 - see the lectures of V. Hussin, J.-P. Derendinger and 
H.P. Nilles in this volume : 
in the latter theories, the supersymmetry generators
relate bosonic and fermionic fields and thus 
represent a symmetry between these fields.

\medskip 

\subsection{Internal symmetries}
\vspace{-9mm}

As an example, we consider the time evolution
of a free particle of mass $m$ and charge $e$  
as described by quantum mechanics. It is governed by the 
{\em Schr\"odinger equation} 
\eq
H \psi = \ri \hbar \, \partial_t \psi 
\ \ ,
\eqn{schr}
where $H = {1 \over 2m} \, \vec{P} ^2 = 
{1 \over 2m} \,({\hbar \over \ri} \, \vec {\nabla} )^2$ 
denotes the Hamiltonian operator and $\psi = \psi (\vec r , t)$ 
the wave function associated to the particle. 
Obviously, equation \equ{schr} is invariant under the  
\underline{global} (i.e. $\vec r$ and $t$ independent)
\underline{phase transformation} 
\eq 
\psi \ \longmapsto \ \psi^{\prime} = {\rm e}^{\ri \alpha} \psi
\ \ ,
\eqn{gpt}
where $\alpha \in {\bf R}$ is a constant\footnote{\small One sometimes
talks about \underline{rigid} rather than global
 transformations so as to avoid topolo\-gical connotations.}.  
Note that $ {\rm e}^{\ri \alpha}$ is an element of the abelian Lie group 
$U(1)$  
of complex numbers of modulus one. 

We emphasize that the invariance \equ{gpt} represents an 
{\em internal symmetry}: it only acts on the space of 
fields (i.e. space-time functions) $\psi$ and not on 
the space-time manifold. 

According to Noether's theorem, the invariance of the 
equation of motion under the continuous symmetry \equ{gpt}
 implies the existence of a conserved 
charge. In fact, one can easily check that the integral  
$\int_{{\bf R}^3} | \psi ( \vec r , t ) |^2 \, d^3x$ 
(which may be associated with the electric charge of the particle)
does {\em not} depend on the variable $t$ if $\psi$ is
 a solution of \equ{schr}.

\medskip 

\subsection{From global to local symmetries}
\vspace{-9mm}

We now consider a \underline{local}
(i.e.  
$\vec r$ and $t$ dependent)
 \underline{phase
 transformation} or so-called 
\underline{gauge transformation} : 
\eq
\fbox{\mbox{$ 
\  \psi^{\prime}(\vec r , t)  = {\rm e}^{\ri \alpha(\vec r, t)}
\,  \psi (\vec r , t)
\ $}}
\qquad {\rm with} \quad  \alpha (\vec r , t) \in {\bf R}
\ \ .
\eqn{gt}
The Schr\"odinger equation \equ{schr} is {\em not} invariant 
under these transformations. To verify this fact explicitly, 
we introduce relativistic notation (and choose a system of 
units in which the velocity of light $c$ equals one) :
 for the derivatives $(\pa_{\mu} \psi ) \equiv (\partial \psi 
 /\partial x^{\mu}) = (\pa_t \psi , \vec{\nabla} \psi )$
 with respect to the space-time 
coordinates   
$(x^{\mu}) _{\mu = 0,..,3} = (t, \vec r \, )$, we have 
$$ 
\partial _{\mu} \psi^{\prime} = {\rm e} ^{\ri \alpha} 
[ \partial _{\mu} \psi + \ri (\partial _{\mu} \alpha ) \psi ]   
$$
and the terms in $\partial_{\mu} \alpha$ do not drop out
of the primed (i.e. gauge transformed) Schr\"odinger equation. 
Thus, if we {\em want} the gauge transformations   
\equ{gt} to be 
a symmetry of the theory, we have to modify the evolution equation 
\equ{schr} in such a way that 
it preserves its form under these transformations. 
For this purpose, we introduce some new fields 
into the equation which transform in a specific way
so as to  ensure its covariance.
The simplest way 
to realize this idea consists of
introducing  a \underline{scalar potential}
$\phi ( \vec r , t) $ and a \underline{vector potential}
$\vec A (\vec r , t)$    
into the free Schr\"odinger equation by means of the 
so-called 
\underline{minimal 
coupling} procedure: one replaces the ordinary derivative by the  
\underline{(gauge-)
covariant derivative}, 
\eq
\partial _{\mu}
\ \longrightarrow \
\ \fbox{\mbox{$ 
\ D^{(A)}_{\mu} \equiv \partial _{\mu} + \ds{\ri e \over \hbar}
\, A_{\mu}   
\ $}}
\qquad {\rm where} \quad  (A_{\mu}) \equiv ( \Phi , - \vec{A}) 
\ \ , 
\eqn{cd}
i.e. explicitly  
$ 
 \partial_t  \rightarrow 
 \partial_t + {\ri e \over \hbar} \Phi ,       
\ \vec{\nabla} \rightarrow   \vec{\nabla} - 
{\ri e \over \hbar} \vec{A}$. 
Furthermore, the \underline{gauge vector field} $(A_{\mu})$
is required to transform inhomogenously under gauge 
transformations,  
\eq
\fbox{\mbox{$ 
\ A_{\mu} ^{\prime} = A_{\mu} - \ds{\hbar \over e} 
\, \partial _{\mu}\alpha
\ $}}
\qquad \Longleftrightarrow \
\ \left\{ \begin{array}{l}
\Phi^{\prime} = \Phi - {\hbar \over e} \, \pa_t \alpha \\
\vec{A} ^{\prime} = \vec{A} + {\hbar \over e} \, 
\vec{\nabla} \alpha
\end{array}
\right.
\ \ , 
\eqn{inhom}
so as to compensate the unwanted terms in $\partial_{\mu}
\alpha$ in the primed equation.
In fact,  
\begin{eqnarray}
( D_{\mu} ^{(A)} \psi )^{\prime} \equiv 
D_{\mu} ^{(A^{\prime})} \psi^{\prime} & = & 
( \pa_{\mu} + \ds{\ri e \over \hbar} \, A_{\mu}^{\prime} ) 
\, ({\rm e}^{\ri \alpha} \psi ) 
\nonumber 
\\
& = & 
{\rm e}^{\ri \alpha} \, \left[ D_{\mu} ^{(A)} \psi + 
\ri (\pa_{\mu} \alpha ) \psi -  
\ri (\pa_{\mu} \alpha ) \psi
\right] 
\ \ , 
\nonumber  
\end{eqnarray}
which means that 		
the covariant derivative of $\psi$ transforms in the 
same way as $\psi$ (whence its name) :  
\eq
\fbox{\mbox{$ 
\ (D_{\mu} ^{(A)} \psi )^{\prime} = 
{\rm e}^{\ri \alpha} \; D_{\mu} ^{(A)} \psi
\ $}}
\ \ .  
\eqn{cod}
After performing the minimal coupling, the free 
Schr\"odinger equation becomes
$$ 
- \ds{\hbar^2 \over 2m} \, ( \vec{\nabla} - 
\ds{\ri e \over \hbar} \,  \vec{A} \, )^2 \psi \, = \, 
\ri \hbar \, ( \pa_t + \ds{\ri e \over \hbar} \, \Phi )
\psi 
$$ 
or 
\eq 
\ds{1 \over 2m} ( \vec{P} - e \vec{A} \, )^2 \psi 
+ e \Phi \, \psi \, = \, 
\ri \hbar \, \pa_t \psi 
\ \ . 
\eqn{schrem}
This differential equation describes 
the coupling of the particle of mass $m$ and charge
$e$ with the \underline{electromagnetic field}
 associated to the 
potential $(A_{\mu})$,  
\begin{eqnarray}
\vec E & \equiv & - \stackrel{\rightarrow}{{\rm grad}}  \Phi 
- \pa_t \vec A 
\\
\vec B & \equiv &   \stackrel{\rightarrow}{{\rm rot}} \vec A 
\ \ . 
\nonumber
\end{eqnarray}
By construction, the interacting Schr\"odinger equation 
\equ{schrem} is invariant under the gauge transformations 
of $\psi$ and $A_{\mu}$ (which leave the field strengths 
$\vec E , \vec B$ invariant).  

Let us summarize once more the whole \underline{gauging 
procedure} in  quantum mechanics : 

\begin{tabular}{|p{16cm}|}
\hline 
\smallskip 
\noindent {\bf (1)}
 The starting point is the Schr\"odinger equation 
for a {\em free} particle : it is invariant under {\em global}
$U(1)$-transformations which implies the conservation of 
electric charge by virtue of Noether's theorem. 

\noindent {\bf (2)} 
The requirement that the Schr\"odinger equation 
 be invariant under {\em local} $U(1)$-transformations 
(and simplicity) implies 
\begin{itemize}
\item the introduction of a gauge field $(A_{\mu})$	
\item that the way in which matter and gauge fields 
{\em interact} with each other is fixed ! 
\end{itemize}
\vspace{-5mm}
\\ 
\hline 
\end{tabular}

\bigskip 

Clearly, the fact that symmetries determine the (form of)
interactions is a very strong result.
The gauging recipe is at the origin of one of the greatest
success stories of theoretical physics : 
the introduction of
\underline{non-abelian gauge (Yang-} \underline{Mills) theories}
led to the development of the standard model of particle 
physics \cite{weinberg,cheng,ryder}. 
After the foregoing discussion of quantum mechanics, 
it is straightforward
 to outline the construction of Yang-Mills theories.  
Instead of the abelian Lie group $U(1)$, one considers 
a non-abelian (and compact) Lie group $G$, e.g. the  
{\em special unitary group}
$$
SU(n) = \{ n\times n \; \mbox{matrices} \ A \ 
\mbox{with complex  
coefficients} \ | \; 
A^{\dagger} A = {\bf 1}_n , \, {\rm det}\, A =1 \}.
$$
Just as it was the case for quantum mechanics, 
the starting point {\em (input)} is the following :
\begin{enumerate}
\item A certain content of {\em matter fields} 
$\varphi_1 (x), ... , \varphi_N (x)$ (scalar 
or spinor fields) which are best assembled into a multiplet 
$\Phi = ( \varphi _1 , ..., \varphi _N )^t$. 
\item A continuous symmetry {\em group} $G$.   
\item A {\em Hamiltonian} (or {\em Lagrangian}) which  
describes the dynamics and self-interaction of the 
fields $\varphi_i$ and   which is {\em invariant} under global 
internal symmetry  transformations  $g \in G$, i.e. under 
$$
g = {\rm e} ^{ \ri \underline{\alpha} }
\qquad {\rm with} \ \; 
\underline{\alpha} = \sum_{a=1}^{{\rm dim}\, G} 
\alpha^a t_a 
\ \ ,
$$  
where 
$\{ t_a \}_{a=1,..,{\rm dim}\, G}$ denotes a basis
of the Lie algebra of $G$ and 
$\alpha^a \in \rr$. (We remark that $g$ acts on the multiplet $\Phi$ 
by means of an $N\times N$ representation matrix $D(g)$.) 
\end{enumerate} 
The gauging procedure ($\alpha^a \to \alpha^a (x)$)
then leads to the introduction of a {\em gauge field}
 with values in 
the Lie algebra of $G$, 
\eq
\underline{A} _{\mu} (x) = \sum_{a=1}^{ {\rm dim}\, G}
A_{\mu} ^a (x) t_a 
\ \ ,
\eqn{ym}
transforming inhomogenously under 
$g(x) = {{\rm e}} ^{{\ri} \underline{\alpha} (x) }$ :  
\eq 
\underline{A} _{\mu} ^{\prime} = g^{-1} \, \underline{A} _{\mu} \, 
g
+ \ds{\ri \over \lambda}  \; g^{-1} \, \partial _{\mu} g 
\qquad (\, \lambda = \mbox{coupling constant} \, ) 
\ .
\eqn{ymt}
(Note that our previous equations are recovered for the case
of the abelian group $U(1)$ for which ${\rm dim} \, G =1, \; 
t_1 = 1, \; \alpha ^1 
\equiv \alpha$ and $\lambda = e / \hbar$.)

Gauge theories based on the groups 
$U(1), \, SU(2)$ and $SU(3)$ describe the electromagnetic, weak 
and strong interactions \cite{weinberg,cheng}
 - see the lectures 
of P. Aurenche in this volume.   
The associated conserved charges are the {\em electric charge}, 
{\em weak isospin} and {\em baryon number}. 

By gauging  translations ($a^{\mu} \to a^{\mu} (x)$),
 one can construct  \underline{general 
relativity} 
which is invariant under the group of 
\underline{general coordinate transformations}
$x \to x^{\prime} (x) \simeq x +a (x)$.  
Similarly, by gauging rigid supersymmetry
 transformations, 
one obtains \underline{supergravity},
i.e. the supersymmetric extension of general relativity 
\cite{sohnius}.  
Actually, general relativity was not constructed 
along these lines at first : Einstein derived his whole theory from 
a single symmetry principle, the \underline{equivalence
 principle} (equivalence of inertial and gravitational mass)
which finds its formal expression in the {\em principle of 
general covariance} (equivalence of all reference frames 
for the description of physical laws) \cite{strau}.
For the interpretation of general relativity as a gauge 
theory, we refer to \cite{sexl}.

\medskip 

\subsection{Combining geometric and internal symmetries}
\vspace{-9mm}

Let us recall 
that $SU(2)$ is a three-dimensional 
Lie group whose Lie algebra $su(2)$ is spanned 
by the Pauli matrices 
$(\sigma_a)_{a=1,2,3}$. 
In field theories admitting $SU(2)$ as an 
internal symmetry group, one can introduce 
fields $\underline{\pi} (x)$ with values in the Lie algebra $su(2)$,
i.e. $\underline{\pi} (x) = 
\sum_{a=1} ^3 \pi^a (x) \sigma_a$. 
For static configurations, the three components 
$\pi^a( \vec r \, )$ can  be related to 
the three components  of the
ordinary space vector $\vec r = (x^a)_{a=1,2,3}$, 
e.g. $\underline{\pi}^a (\vec r \, ) = f (r ) x^a$ 
where $r = \sqrt{(x^1) ^2 + (x^2) ^2 + (x^3) ^2}$, 
as in the 
\underline{Skyrme model} of nuclear physics \cite{skyrme}.   
Simi\-lar  examples are given by the \underline{monopole}
and \underline{instanton} configurations which occur 
in $SU(2)$ gauge theories \cite{ryder}. Thus,   
for these configurations (states), 
there is a certain mixing 
of indices associated with internal and geometric 
invariances  
(which are - a priori - completely unrelated). 

One may wonder whether such a blend of internal and geometric 
symmetries may exist at a more fundamental level 
as a general feature of field theory and not simply 
 in specific field configurations 
of particular models. 
(This feature would be very attractive for the construction of 
a unified theory of all fundamental interactions
 including gravity.) 
That this is {\em not} possible is expressed by the 
so-called \underline{no-go theorems}, in particular 
the theorem of Coleman and Mandula,
which essentially says the following :  
the most general invariance  group 
of a relativistic quantum field theory is a direct product 
of the Poincar\'e group and an internal symmetry group, 
i.e. there is no mixture of these symmetry transformations. 

However, these no-go theorems do 
{\em not} claim that such a mixture 
cannot exist if the set of all symmetry transformations
represents a more general algebraic structure 
than a {\em Lie group}. Indeed, 
a famous result known as the        
\underline{theorem of Haag, Lopusz\`anski and Sohnius} \cite{sohnius}
states that the most general {\em super} Lie group of 
symmetries of a local field theory is the 
$N$-{\em extended super Poincar\'e group} in which there 
is a non-trivial mixing of geometric transformations 
and internal $SU(N)$ transformations.
As a matter of fact, this result can also be viewed
as a good argument in favor of the existence of 
supersymmetry as an invariance of nature since it states that  
{\em supersymmetry is the natural} (only possible) 
{\em symmetry} if one allows for super Lie groups as 
symmetry structures.       

\medskip 

\subsection{Duality symmetries}
\vspace{-9mm}

Let us consider {\em Maxwell's equations} with both electric 
and magnetic sources 
(i.e. hypothetical magnetic monopoles) :
\begin{eqnarray}
\nonumber 
{\rm div} \, \vec E & = & \rho_e 
\quad \, , \quad 
\stackrel{\rightarrow}{{\rm rot}}  \vec E + \partial_t \vec
B \, = \, - \vec{\jmath} _m 
\\
{\rm div} \, \vec B & = & \rho_m 
\quad , \quad 
\stackrel{\rightarrow}{{\rm rot}}  \vec B -  \partial_t \vec
E \, = \, \  \vec{\jmath} _e 
\ \ .
\end{eqnarray}
These equations are invariant under the \underline{duality 
transformation}
$$
(\vec E, \vec B,
\rho_e, \rho_m, 
\vec{\jmath}_e, \vec{\jmath}_m ) \longrightarrow
(\vec B, - \vec E, \rho_m, - \rho_e,  
\vec{\jmath}_m   - \vec{\jmath}_e)
\ \ .
$$ 
In quantum mechanics, the presence of monopoles 
leads to the quantization of both electric  and magnetic 
charges :  for a given 
particle, these charges have to be related by 
Dirac's condition 
$q_e q_m = 2 \pi n$ where $n$ denotes an entire number. 
Thus, the minimal charges obey 
$q_m = {2 \pi \over q_e}$. Since the duality symmetry 
exchanges electric and magnetic variables, we conclude from this 
relation that  
{\em duality exchanges the coupling constant $q_e$ with its inverse} (up to the 
factor $2 \pi$) or that

\begin{tabular}{|p{15cm}|}
\hline  
Duality symmetries exchange weak and strong 
coupling regimes. 
\\
\hline
\end{tabular} 

\noindent 
Henceforth, it is possible to learn 
about strong-coupling physics from the weak-coupling physics 
of a dual formulation of the theory
\cite{stefan}.
Following a seminal paper by N. Seiberg and E. Witten, 
the latter  idea proved to be extremely useful in recent years 
in the context of both field and string theories. 
It led to a further  study of other extended objects like 
\underline{membranes} or even higher-dimensional 
objects, all of which are referred to as \underline{$p$-branes}
($p=1$ for strings, $p=2$ for two-dimensional membranes,...).  
All of these objects and theories 
 seem to be related in amazing ways by 
duality transformations.

\newpage 

\subsection{Miscellaneous}
\vspace{-9mm}

The fact that the rescalings $x \to {\rm e}^{\lambda} x$ do
 {\em not} represent an invariance of nature was already 
elucidated by G. Galilei \cite{feynman}. Yet, this 
notion plays an important r\^ole in natural 
phenomena : in many places,
one encounters  \underline{self-similar
structures}, i.e. structures looking exactly the same way at 
different scales and \underline{fractal objects}, 
i.e. objects which are self-similar up to statistical
 fluctuations \cite{siva,schroeder}.

The discrete geometric symmetries of parity (P) and 
time reversal (T) are usually discussed in conjunction with 
\underline{charge conjugation} (C) which transforms a 
charged particle into its anti-particle with the  
opposite charge. This represents an internal symmetry 
relating a complex field to its complex conjugate.
The celebrated \underline{PCT theorem} states that 
the product of all these discrete symmetries 
is conserved 
in any local quantum field theory \cite{weinberg}. 

Elementary particles are not only characterized
 by their {\em mass}
and {\em spin} (related to geometric symmetries) 
and their {\em charge} with respect to gauge symmetries
(e.g. electric charge and weak isospin for the $U(1)$- and 
$SU(2)$-invariances),
 but also by other {\em additive
 quantum numbers} like the leptonic numbers. 
These can always be associated with a global $U(1)$ symmetry 
group since $g(Q_1)  g(Q_2) = g(Q_1 + Q_2)$ for 
$g(Q) = {\rm exp}\, ({\ri} Q \alpha) \in U(1)$. 
Thus, the corresponding conservation laws can also 
be traced back to an 
$U(1)$-invariance of the Lagrangian.

\section{The mathematical description of symmetries
and their implementation in physical theories} 

\vspace{-2mm}

\subsection{Mathematical description : (Lie) groups and algebras}
\vspace{-9mm}

By studying the composition of symmetry 
transformations, e.g. of geometric objects, 
one reaches the conclusion that they form a {\em group} 
and, more specifically, a {\em Lie transformation group} if one 
considers 
continuous, finite symmetry transformations. (One talks 
about {\em transformation} groups if the group elements 
operate on a certain space.) 

Let us mention a few prototypes of \underline{groups}. 
The sets ${\bf Z}$ and ${\bf R}$ of entire and real numbers,
supplemented with the  law of addition as `group multiplication'
are examples of abelian (i.e. commutative) groups. 
A very useful group can be
 constructed from the additive group  
$({\bf Z},+)$ 
by identifying  even and odd numbers, respectively :
 this is the 
{\em quotient (factor) group}  
$$
{\bf Z}_2 \equiv {\bf Z} \, / \, 2{\bf Z} = \{ 
\overline 0 , \overline 1 \} 
\ \ ,
$$
 which consists of equivalence classes of even and
 odd numbers.
 The `multiplication
table' reads
$$
\overline 0 + \overline 0 = \overline 0
\qquad , \qquad 
\overline 0 + \overline 1 = \overline 1 = 
\overline 1 + \overline 0
\qquad , \qquad 
\overline 1 + \overline 1
= \overline 0 
\ \ . 
$$
In physical applications, the dichotomy $\overline 0 /
\overline 1$ has different 
interpretations depending on the field considered : even/odd,  
spin up/spin down,   
 integer spin/half-integer spin (i.e. boson/fermion field), ...
Concerning the latter interpretation, we note that 
the multiplication table of the group ${\bf Z}_2$ 
reflects the general 
structure of the 
commutation relations of ${\bf Z}_2$-graded (`super')
 Lie algebras - 
see equations \equ{schema}.

The examples considered so far already convey an idea 
of the variety of  groups one can 
encounter : ${\bf Z}_2$ is a {\em finite discrete} 
group, {\bf Z} is an {\em infinite discrete} group 
whereas ${\bf R}$ is an infinite group characterized by 
one continuous real parameter : this is an example of a 
(one-dimensional) Lie group. Roughly speaking, 
a \underline{Lie group} is a group whose 
elements can be   
parametrized by one or several real numbers (the total 
number of 
which is referred to as the {\em dimension} of 
the Lie group). Thus, $({\bf R}^n,+)$ is an
$n$-dimensional, abelian Lie group. 

Before pointing out further examples and applications, 
it is useful to recall that a given group admits    
many disguises : two groups are \underline{isomorphic} 
to one another
if there is a one-to-one correspondence between their elements
 and if 
they have exactly the {\em same structure}.
For instance, 
the group $({\cal T}_n, \circ )$ of translations of ${\bf R}^n$
(introduced in section 3) 
is isomorphic to the additive group $(\rr^n , +)$, because
the correspondence $T_a \leftrightarrow a$ is one-to-one 
and preserves the group `multiplication' :
$T_{a} \circ T_{b} = T_{a + b}$ . 

There is a large number of finite-dimensional Lie groups 
of physical importance.
The examples we encountered 
so far include the groups of translations and rotations, the 
groups of Lorentz, Poincar\'e and conformal transformations   
(which are all related to geometric 
symmetries) or the special unitary groups 
 related to internal symmetries. 
The latter (and various others) also occur as 
{\em classification groups} in  
atomic, nuclear, particle and solid state  physics.    
 
Apart from these finite-dimensional Lie groups, some  
infinite-dimensional ones, for which we will give three 
examples, play an important r\^ole in physics.  
The  Lie group of {\em diffeomorphisms} of the space-time 
manifold (`reparametrizations'
or `general coordinate transformations') is at 
the very heart of general relativity. The {\em Virasoro group}, 
i.e. the group of diffeomorphisms of the unit circle,   
is fundamental in two-dimensional conformal field theory and its 
applications to statistical mechanics   
or solid state physics \cite{mack}. Finally, the group of 
{\em area-preserving diffeomorphisms} of a (hyper)surface 
manifests itself, amongst others, in the quantum Hall effect
 - see the lectures of A. Cappelli in this vo\-lume.

Instead of a Lie group, one often considers the associated 
\underline{Lie algebra} : this is  the set of 
infinitesimal transformations, supplemented with a
{\em Lie bracket} 
which is the ordinary commutator $[A,B] = AB-BA$ in the 
case of matrix  algebras. 
By exponentiating the Lie {\em algebra} elements 
(i.e. integrating up the infinitesimal transformations),  
one recovers
the Lie {\em group} elements.

\medskip 

\subsection{Physical implementation : representations}
\vspace{-9mm}

In physics, one is often confronted with the 
problem of letting a given symmetry transformation  
(e.g. a translation or rotation in three-dimensional space)
 act on other objects 
(e.g. the wave function describing an electron
 in quantum mechanics or 
the vector field describing a force). 
The appropriate mathema\-tical tool for achieving this goal   
is the one of a  
 {\em representation} of the group (or Lie group 
or Lie algebra, depending on the type of symmetries  
considered). For instance, an $N$-dimensional  
\underline{representation $D$ of the group $G$}
 is defined as follows : 
 to each element $g\in G$,
 one associates an $N\times N$-{\em matrix} $D(g)$ 
(i.e. a {\em linear operator}  
on an $N$-dimensional vector space) such 
that {\em the group structure is preserved,} i.e. 
$D(g g^{\prime} ) = D(g) D(g^{\prime})$. If the 
correspondence $g \mapsto D(g)$ is one-to-one,    
the set of all representatives $D(g)$ forms a group 
which is isomorphic to the original group $G$.
E.g. an (infinite-dimensional) representation of the
 translation group 
$({\cal T}_1 , \circ )$ on the Hilbert space of 
 wave functions $\psi (x)$  is defined by 
 $T_a \mapsto D(T_a)$ with 
$D(T_a) \psi \equiv \psi_a$ and $\psi_a (x) :=\psi (x-a)$.

\medskip 

\subsection{Generalizations}
\vspace{-9mm}
 
Besides  (Lie) groups and algebras, various 
other algebraic structures have been invoked in physics 
over the last decades for the description of  
symmetries. Let us mention a few of them.

The renormalization transformations occurring in the theory 
of dynamical systems \cite{ce}, in statistical mechanics
 or quantum field theory
\cite{sewell} are not inver\-tible in general and
 therefore only form a 
{\em semigroup}, the \underline{renormalization
semi-} \underline{group}, traditionally referred to as renormalization
{\em group}. 

\underline{Lie superalgebras} or \underline{${\bf Z}_2$-graded 
algebras} (whose definition was outlined in section 3)  
are ${\bf Z}_2${\em -graded extensions of ordinary
 Lie algebras}.  
As we pointed out, 
the  super {\em Poincar\'e} algebra represents 
the basis of all supersymmetric field theories.   
${\bf Z}_2$-graded extensions of other Lie algebras  
like $su(n)$ also admit numerous physical applications
\cite{corwin}.

For a Lie algebra, the commutator of two ele\-ments
is again an algebra element, i.e.  
a linear combination of basis elements of the algebra. 
As first shown by A.B. Zamolodchikov, one can construct in a 
consistent way algebras for which the commutator
involves products of algebra elements : these
{\em non-linear generalizations of Lie algebras} are known as
\underline{$W$-algebras}. They occur in numerous 
contexts, e.g. as symmetry algebra
of the hydrogen atom and other elementary quantum systems 
\cite{tjin} or as invariances of field theoretical models
 \cite{schoutens}. 

The so-called \underline{quantum groups} (i.e. 
{\em deformations
of groups})   
as well as related 
algebraic structures like {\em quasi-triangular Hopf algebras}
are currently   
studied and applied  in various ways to  
quantum physics \cite{mack, quantum}.    

The {\em current algebras} 
of two-dimensional field theory 
 represent examples of affine \underline{Kac-Moody algebras},  
which are to be  combined with \underline{Virasoro algebras}
in conformally invariant models 
\cite{go}. 

In the remainder of this section, we will 
describe in some detail the so-called \underline{BRS
 (Becchi-Rouet-Stora) 
algebra} for Yang-Mills theories with structure
group $G$  \cite{gieres,bertl}. This algebra 
allows to express the gauge invariance at the
 quantum level of the theory and
thus to impose restrictions on the possible counterterms 
that can be added to the quantum action. 
Henceforth, the BRS symmetry is the clue to the 
proof of {\em renormalizability} \cite{piguet} - 
see the lectures of O. Piguet in this volume.   
It is also an essential tool for the algebraic
characterization and  determination of {\em anomalies} 
occurring in quantum theory - see 
section 6. 

To start with, we consider the infinitesimal form of 
the gauge transformation \equ{ymt} :
with $g = {\rm e}^{\ri \underline{\alpha}} \simeq  
{\rm 1} + \ri \underline{\alpha}$, 
we have 
$\underline{A} _{\mu} ^{\prime} \simeq \underline{A} _{\mu} 
- {1 \over \lambda} \,  
\delta _{ \underline{\alpha}} \underline{A} _{\mu} $ with 
$$
\delta _{ \underline{\alpha}} \underline{A} _{\mu} 
 = D_{\mu} \underline{\alpha}
\equiv \partial _{\mu} \underline{\alpha} - {\ri} \lambda 
 \, [ \underline{A}_{\mu} , \underline{\alpha} ]  
\ \ ,
$$ 
i.e. the gauge covariant derivative of the Lie algebra-valued 
parameter $\underline{\alpha}$. 
The commutation relation of these transformations reads 
\eq
[ \delta _{ \underline{\alpha}} \; , 
\delta _{ \underline{\alpha} ^{\prime} } ] = 
- {\ri} \lambda 
\, \delta _{ [ \underline{\alpha} \, ,
\, \underline{\alpha} ^{\prime} ]}  
\ \ . 
\eqn{ymit}
By virtue of this relation, 
the set of infinitesimal gauge transformations 
$\delta _{ \underline{\alpha}}$ is 
a {\em representation} of the 
\underline{gauge algebra}
${\cal G} = \{ \, \underline{\alpha} : 
\rr^4 \to {\rm Lie} \, G \, \}$
where ${\rm Lie} \, G$ denotes the Lie algebra of $G$. 

In order to obtain the BRS algebra of this theory, 
 the parameter
 $\underline{\alpha}$ is turned into a {\em Faddeev-Popov 
ghost} field $\underline{c}$, i.e. a relativistic 
field with {\em `ghost 
number'} one. All classical fields like $\underline{A} _{\mu}$ 
or the matter fields are assigned
 the ghost number zero. 
(Note that the introduction of the ghost number in the algebra 
of fields amounts to the definition of a 
{\em grading} by ${\bf Z}_2$.) We now introduce the 
\underline{BRS-operator $s$} by the following requirements.  
The $s$-variation 
of the classical field $\underline{A}_{\mu}$ 
is an infinitesimal gauge transformation 
with  $\underline{\alpha}$ replaced by
 the field $\underline{c}$. The operator $s$ is linear 
and acts on products according to 
$s(PQ) = (sP) Q + (-1)^{g(P)} PsQ$ where $g(P)$ denotes the 
ghost number of $P$. The $s$-variation of the ghost field 
$\underline{c}$ 
is defined in such a way that $s$ is 
{\em nilpotent} (i.e. $s^2 = 0$)
 on all fields. This implies
\eq 
\fbox{\mbox{$
s \underline{A} _{\mu}  =  D_{\mu} \underline{c}
\quad , \quad 
s \underline{c}  =  \ds{{\ri} \lambda \over 2} \, [ \, \underline{c} \, ,
\, \underline{c} \, \} 
= {\ri} \lambda \, \underline{c} \, \underline{c}
$}}
\ \ ,   
\eqn{del}
where $[\underline{P}, \underline{P} ^{\prime} \}= \underline{P}
\, \underline{P} ^{\prime} - (-1)^{g(\underline{P})g(\underline{P} ^{\prime})}
\underline{P} ^{\prime} \, \underline{P}$ represents the   
graded commutator of the Lie algebra elements
$\underline{P}$ and $\underline{P} ^{\prime}$. 
We remark that  the BRS-variation of the ghost field amounts to 
``reading the other way round" the commutation 
relations \equ{ymit} and that $s^2 \underline{c} = 0$ 
reflects the Jacobi identity of the Lie algebra ${\cal G}$.
Since $s$ raises the ghost number by one unit and has 
properties   that are analogous to the exterior derivative $d$
of differential forms, the BRS-algebra may be viewed as  
 the 
{\underline{differential algebra} 
associated to the Lie algebra of infinitesimal gauge
transformations.

\section{Implications of symmetries for
the formalism, the results and the structure of physical 
theories}

In the following, we will try to gather the conclusions 
drawn from the examples studied in the 
previous sections and to highlight the salient features 
and implications of symmetries in physics. 
Probably the latter are best summarized by the following 
observation of H. Weyl \cite{weyl} : {\em ``As far as I see, 
all a priori statements in physics 
have their origin in symmetry".} 

\begin{tabular}{|p{16cm}|}
\hline 
\smallskip 
The laws of nature are possible realizations of the symmetries
of nature. 

\medskip  
The basic quantities or {\em building blocks} of physical 
theories 
are often defined and classified 
by virtue of symmetry considerations,
 e.g. elementary particles and relativistic fields.

\medskip 
The {\em general structure} of physical theories is largely 
determined by the underlying invariances. In particular, 
the {\em form of interactions} is strongly res\-tricted 
by geometric symmetries (relativistic covariance)
and the gauge symmetries essentially fix all fundamental 
interactions (electro-weak, strong and gravitational forces).
\bigskip 
\\ 
\hline 
\end{tabular}

\noindent

From a practical point of view, one can say : 

\begin{tabular}{|p{16cm}|}
\hline 
\vspace{-2mm}
\begin{itemize}
\item The knowledge and study of the symmetries of nature provide 
deeper insights into the properties of physical systems. 
\item The exploitation of symmetries often simplifies 
the determination of solutions  
of a physical problem and, in some cases,  
a complete or exact solution only becomes accessible if  
a sufficient amount of symmetry is present.   
\end{itemize} 
\vspace{-6mm}
\\ 
\hline 
\end{tabular}

\bigskip
As an illustration of the first of these points, we recall  
how the characteristics of space-time affect physics : 

\begin{tabular}{|p{16cm}|}
\hline 
Properties of space and time 
$ \ \ \Longrightarrow \ \ $
Symmetries (invariances) of equations  
$ \ \ \stackrel{{\rm Noether}}{\Longrightarrow} \ \ $ 
Conservation of energy, momentum and angular momentum
\\ 
\hline 
\end{tabular}

\bigskip 

\noindent Even if one doesn't know anything 
about the symmetries of physical laws or 
if one doesn't take them into account when solving 
a concrete problem, they manifest themselves in the 
solutions or results. For instance, if one is not aware of  
the symmetries of Newton's equation 
$m \ddot x  = - {dV \over dx}(x)$, one finds that, for its
 solutions $x = x(t)$, the combination 
${1\over 2} \, m \dot x (t)^2 + V(x(t))$ does {\em not} depend 
on the variable $t$. Of course, this quantity is nothing 
but the {\em energy} whose conservation is due to the {\em time
translation invariance} of the equation of motion.

Illustrations of the second point are given by the study of 
{\em time evolution equations} of classical and quantum physics. 
Many of these differential equations  
are expressions of invariants associated to some Lie group 
and the theory of these groups provides a unifying 
viewpoint for the study 
of {\em all} special functions  
and {\em all} their properties \cite{satt,vilenkin}. 
In fact, S. Lie invented the theory of Lie groups when 
studying the symmetries  of 
 differential equations. 
The integration of a (partial) differential equation 
by the method of {\em separation 
of variables} \cite{miller} 
or by {\em Lie algebraic methods} is
 intimately connected with 
 the existence 
of symmetries  \cite{olver, adams}.
In particular, the exact solubility of the 
Schr\"odinger equation 
in quantum mechanics can be traced back to the 
underlying invariances \cite{iach}.
In the latter case, the investigation of symmetries allows 
for an interpretation of the {\em degeneracies}
which generally occur in
 the energy 
spectrum of quantum systems - see the lectures of 
M. Kibler in this volume.  

In many non-linear field theories 
like general relativity or Yang-Mills
theories,   
the basic field equations 
are highly non-linear and exact solutions
are  only known for `sufficiently symmetric' distributions 
of matter
(e.g. rotationally or axially symmetric configurations)
 \cite{exact}.

\section{Different manifestations of symmetries}  

\vspace{-2mm}

So far we have discussed {\em exact}  symmetries of the laws 
of nature without  inqui\-ring  about their domain 
of validity. When dealing with such symmetries, 
it is natural to ask the following questions :

\begin{tabular}{|p{16cm}|}
\hline 
\begin{itemize}
\item Symmetries {\em of what} ? 
You may either have symmetries of the {\em equations of 
motion} (the Lagrangian or Hamiltonian) and of the boundary conditions
or symmetries of the {\em solutions} (the states). 
\item Symmetries {\em at which scale} ? At the microscopic or 
macroscopic scale, at low or high energies, temperatures, ... ?
\item Symmetries {\em at what level} ? 
Symmetries may occur in the {\em classical theory} or 
in {\em quantum theory}. 
\item {\em What type} of symmetry ? A symmetry that is 
{\em exact, approximate} or {\em broken} ?
And if the latter applies : do you have an  
{\em explicit, spontaneous} or {\em anomalous breaking}
 of symmetry ?
\end{itemize}
\\ 
\hline 
\end{tabular}

\noindent Obviously, some of these issues
 are related to some others. 
While focusing on the last point, we will touch upon 
all the others. 

\medskip 

\subsection{Broken symmetries}
\vspace{-9mm}

\noindent 
{\bf Explicit breaking}

When an atom is placed in an electric or magnetic 
field, its rotational symmetry gets broken
 (Stark and Zeemann effects). 
The Hamiltonian then involves an additional contribution 
which is {\em not} invariant under rotations : this is referred
 to as an \underline{explicit (`effective') breaking
 of symmetry}.

A similar situation is encountered in field theory if one 
considers a  
Lagrangian which is invariant under certain symmetry 
transformations and then adds a term which does not share 
this invariance 
(e.g. mass terms in the free $SU(3)$ quark
 model \cite{cheng}). 

Even in the case where a symmetry is explicitly broken,
 important conclusions 
can be drawn from this symmetry. For instance,  
if the breaking term has a small amplitude, {\em perturbation 
theory}  around the solutions of the symmetric theory 
can be applied.
In field theory, the {\em conserved} charges of the symmetric 
theory are {\em time-dependent} in the presence of a symmetry 
breaking term, but they still fulfill the same 
{\em current algebra,} from which far reaching conclusions 
can be inferred \cite{cheng}.

\smallskip

\noindent
{\bf Anomalous breaking}

If a certain number of symmetries is present in the 
{\em classical} theory and not all of these symmetries
exist in the corresponding {\em quantum} theory, then 
one talks about an 
\underline{anomalous breaking of symmetries} \cite{ bertl}.
The term expressing the non-invariance of the effective 
action in quantum theory is referred to as \underline{anomaly}. 
J. Wess and B. Zumino pointed out that 
this term 
can be characterized by an algebraic consistency condition. 
Later on, it was shown  by R. Stora that 
the latter equation for the anomaly ${\cal A}$ 
can be rewritten in terms of the BRS-operator $s$ 
as $s {\cal A} = 0$ and that solutions of it  
can be constructed by applying
 the BRS symmetry algebra of the 
theory \cite{bertl, gieres}.  

The presence of anomalies in quantum field theories spoils 
their  renorma\-lizability. Therefore, their {\em absence} in the 
standard model of particle physics is crucial : it is 
equivalent to the {\em equality} of the number of quark families 
and lepton families. In the framework of string theories,
the requirement that anomalies do not occur, leads to
 the conclusion that these strings can only propagate 
in space-time manifolds of a specific dimension, 
the so-called {\em critical} dimension ($d=10$ for  
superstrings).    
Let us also mention that 
the so-called {\em axial anomaly} of field theory 
 manifests itself in observable physical processes,
e.g. the decay of neutral pions 
into photons \cite{jackiw}. 

\smallskip 

\noindent
{\bf Spontaneous breaking}

One talks about a \underline{spontaneously broken
(or hidden) symmetry} if 
a symmetry of the  equations of motion
and boundary conditions is not present in the observed 
solution or, more specially,   
if the quantum {\em states} 
(in particular the fundamental state or vacuum 
in quantum field theory) have less symmetry than the 
{\em equations of motion} (the Hamiltonian or Lagrangian)
\cite{jan, weinberg}. 

A simple non-relativistic model with spontaneous symmetry 
breaking is given by an \underline{infinite
 ferromagnetic substance} : 
at each site $\vec x \in {\bf Z}^3$ 
of a regular cubic lattice,  there
is a spin $\vec S (\vec x \, )$ and the magnetic interaction 
of neighboring spins contributes a term  
$- \vec S (\vec x \, ) \cdot \vec S (\vec y \, )$   
to the energy. Thus, the {\em Hamiltonian},
 which is obtained by  
summing up all these contributions over the lattice, 
is invariant under a   
rotation of all spins by the same amount. 
The {\em fundamental state} is the one for which the energy 
is minimal, i.e. the one for which 
all spins are parallel : clearly, 
this state is {\em not} invariant under the symmetries 
of the Hamiltonian since application of a 
rotation transforms it into another  state (which is physically 
equivalent to the first one).
Similar examples can be found in many systems \cite{chossat},
e.g. in nuclear and 
condensed matter physics \cite{joliot}. 

Rather than limiting oneself to the observation that  
the phenomenon of spontaneous 
symmetry breaking occurs in nature,  one 
 can apply this idea to the 
construction of physical models, e.g. of \underline{supersymmetric 
field theories}. In fact, if  
 supersymmetry is realized as an exact and  fundamental
 symmetry of nature, 
then for each known particle of integer/half-integer spin,
 there ought to be 
a `supersymmetric partner' of half-integer/integer spin and 
of exactly the same mass. However, particles with the latter
properties 
are not observed. This rules out supersymmetry
as an {\em exact} invariance of nature, but not as a 
{\em spontaneously broken} symmetry : in the latter
 case, the invariance exists, 
but the states of the theory 
do not share it, which means that  the 
 superpartners of the known particles actually have greater
 masses than the known  ones.
The mechanism of spontaneous symmetry breaking is 
applied in a similar  vein in the \underline{standard model}
 of particle 
physics to `give' masses to the gauge bosons which  
mediate short-range interactions.        
  
In conclusion, we want to present  another nice 
manifestation of hidden symmetries, namely   
\underline{turbulence} \cite{frisch}.   
This phenomenon also illustrates the fact
that {\em different symmetries may occur at different scales} 
given  by a certain {\em order parameter} of the theory.   
Let us consider the uniform flow of a viscous fluid 
around an infinitely long cylinder. 
For this system, the only order 
parameter is the {\em Reynolds number} 
$R = L V / \nu$ where $L$ is the diameter 
of the cylinder, $V$ the velocity
of the flow and $\nu$ its viscosity. . 
This physical system is described by the {\em Navier-Stokes
equations} supplemented with appropriate boundary conditions. 
We assume that the flow is incident from the left. 
For small values of the Reynolds number, one observes a 
maximal 
number of symmetries : 
\begin{itemize}
\item Left-right symmetry 
\item Up-down symmetry
\item Time translation ($t$-) invariance
\item Invariance with respect to translations along the 
direction of the cylinder.  
\end{itemize} 
As the Reynolds number is increased,
these symmetries disappear one after  the other : 

\begin{figure}[h!]
\centerline{\includegraphics*[height=13.5cm,angle=0]{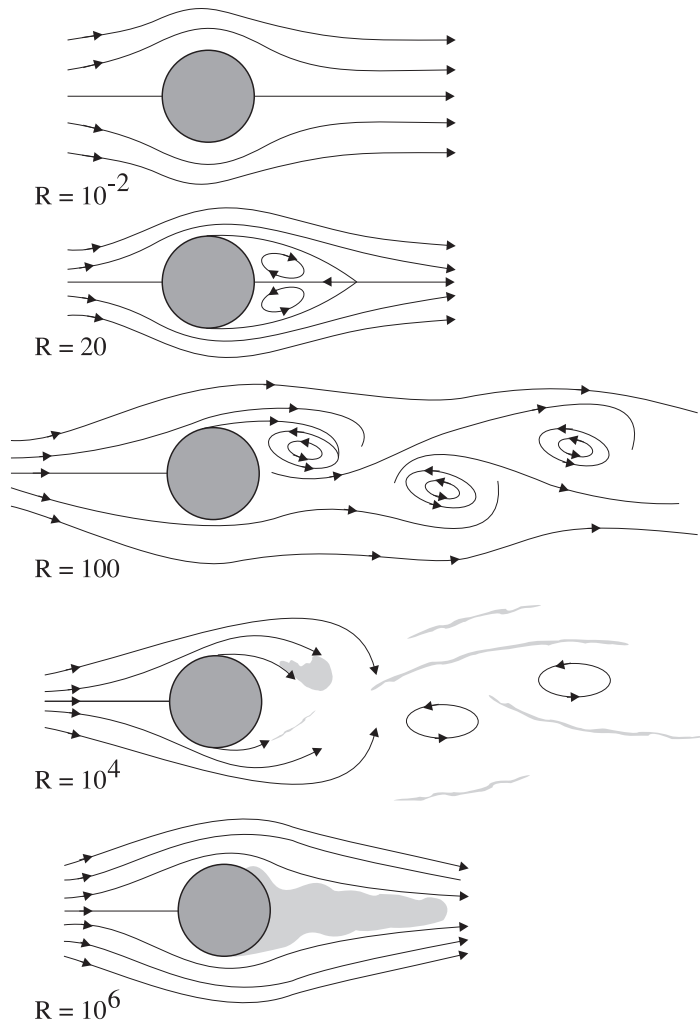}}
\caption{drawn following Feynman's lectures \cite{fl}}
\end{figure}

\newpage
First,  recirculating eddies form behind the cylinder 
and detach themselves in a periodic way, giving rise 
to the K\'arm\'an street of alternating vortices. 
Thus, the first two symmetries are broken  
 and a {\em discrete}
$t$-invariance is all that is left over  
from the {\em continuous} $t$-invariance. 
Upon further increase of the order parameter $R$, the last
invariance also breaks down.      
This represents a spontaneous breaking  of symmetries, 
because the symmetries of the  equations of motion
and boundary conditions are not present in the observed 
solution for large Reynolds number.  

Interestingly enough, 
in the present example, all symmetries are again {\em res\-tored} 
 in a statistical sense (and far from the boundaries)
for  very large values of 
$R$ :  in fact, a {\em homogeneous isotropic 
turbulence} is observed in this case.  
It should be noted that 
{\em complete} \underline{symmetry in statistical physics} means  
a {\em completely irregular} distribution when 
space- or time averages are considered.  (There are 
no spatial correlations 
in the distribution
- see \cite{frisch}  
for more precise definitions.)

\medskip 

\subsection{Miscellaneous and some important asymmetries}
\vspace{-9mm}

We conclude with some remarks on the breaking of
the basic geometric symmetries 
in physical systems. 
{\em Lorentz invariance} is  broken in finite temperature 
field theory since the heat bath defines a privileged 
reference frame \cite{lebel}
- see the lectures of C. Lucchesi.
Some of the {\em discrete symmetries} 
($P,C$ and $T$) or combinations thereof
 are violated in the microscopic world \cite{cecilia}.    
Actually, some of them 
 are also broken at the macroscopic 
level, in particular in {\em thermodyna\-mics} : 
while the basic equations of classical or quantum mechanics 
governing the dynamics of particles are invariant 
under the operation of time reversal (i.e. describe
\underline{reversible processes}), the processes 
observed at the macroscopic level, like the conduction of heat, 
do not have this property and are {\em irreversible}
\cite{balian}. 
A further example is provided by 
{\em cosmology}, i.e. the study of the evolution of 
the universe. Note that the latter does not represent 
a physical system 
like any other since it is {\em unique} ; it 
 is believed to have had a definite time of beginning 
and it is  
expanding. Therefore,  {\em cosmological time}
will not look the same if delayed 
in time, i.e. things are not invariant 
under translations in time.   
All these puzzles related to the \underline{arrow of time}
have intrigued scientists 
since Boltzmann's pioneering work on statistical mechanics 
and they have stimulated 
intensive research \cite{davies}. 

Mention should also be made of two other asymmetries
to whose understanding a lot of endeavor and work 
has been devoted in recent years. These are the    
\underline{baryon asymmetry in the universe}
(i.e. the question why matter is much more abundant in our 
universe than antimatter) \cite{turner}
and the origin of \underline{chiral asymmetry in living systems}
(i.e. the fact that some of the basic organic 
 structures  only occur with one chirality in nature) 
\cite{chiral}.

\section{Conclusion}
\vspace{-2mm}

The reader may find it disappointing or unsatisfactory to 
close our excursion on symmetries with {\em asymmetries}. 
However, symmetries and asymmetries often coexist in nature
just as bad and good moments are parts of our daily life :
they all have their right of being and without the hard times
you would probably not appreciate the better ones
to the right extent.

\bigskip 
\bigskip 
\medskip 

\noindent
{\bf Acknowledgments}
\smallskip

I wish to thank C. Doyen, S. Gourmelen, M. Kibler,
 C. Lucchesi and S. Theisen for pleasant discussions 
and useful remarks on the symmetries of nature.  
I also express my gratitude to Z. Hernaus for his skillful
production of the graphics.

\newpage 

\appendix 
\section{Appendix} 

\vspace{-2mm}

Further details concerning the notions summarized in this appendix
are to be found 
in references \cite{satt}. 

\smallskip 

\subsection{On (Lie) groups and algebras}

\vspace{-12mm}

\begin{defin}
A \underline{group} $(G,\circ )$ consists of a set $G$ together 
with a composition law denoted by $\circ $
which associates an element $x\circ y \in G$ to each pair of 
elements $(x , y) \in G\times G$ such that the following 
properties are satisfied :
\begin{enumerate}
\item Associativity : $\ x \circ (y \circ z) = (x\circ y) \circ z  
\;$   for all $\; x,y,z \in G$.
\item There exists a neutral element (identity), i.e. an element 
$e \in G$ which is  
 such that $\ e \circ x = x \circ e  = x \ $ for all $x \in G$. 
\item For each $x \in G$, there exists an inverse,
 i.e. an element 
 $x^{-1} \in G$ 
such that  
$x \circ x^{-1} = x^{-1} \circ x = e$. 
\end{enumerate}   

The group is said to be \underline{abelian} if 
the commutative law $x\circ y = y \circ x$
 holds for all 
$x,y \in G$. 

If the elements of $G$ only satisfy the first two properties, 
then $(G,\circ )$ is called a \underline{semigroup}. 

\end{defin}

If the group operation is `multiplicative' (e.g. 
multiplication of real numbers), then the identity is denoted 
by $1$. If the operation is `additive' (e.g. addition of real numbers),
 the identity is denoted 
by $0$ and the inverse $x^{-1}$ by $-x$. 

Examples of (abelian) groups are given by 
$(\rr , +)$, $({\bf Z}, +)$ and $({\bf R}-\{0\}, \cdot)$, 
$(\rr_+, \cdot )$ where $\rr_+$ denotes the set of  
positive real numbers
which may be identified with 
 $\{ {\rm e}^a \, | \, a \in {\bf R} \}$.
On the other hand, $({\bf Z}-\{0\}, \cdot)$ 
is only a semigroup,
because the inverse $1/n$ of an integer $n$ is not an integer
(except for  $n=\pm 1$).  

Some particularly interesting groups are the 
\underline{Lie groups}, i.e. groups whose 
elements can be   
parametrized by one or several real numbers ;
the number 
of these parameters is referred to as the {\em dimension} of 
the Lie group\footnote{\small More precisely \cite{broe},
 a Lie group is a 
smooth manifold $G$ which is also a group such that 
the group multiplication and the map sending $x$ to $x^{-1}$
are smooth maps.}. 
For instance, $({\bf R}^n, +)$ is an
$n$-dimensional abelian Lie group and so is the group 
$({\cal T}_n , \circ )$ of translations of 
${\bf R}^n$ introduced in section 3.1 $({\cal T}_n 
= \{ T_a \, | \, a \in \rr ^n \})$. 
$({\bf R}-\{0\}, \cdot)$ and  
$(\rr_+, \cdot )$
are $1$-dimensional (hence abelian) Lie groups.  

An important example of a non-abelian Lie group is given 
by the  {\em special orthogonal group}
$SO(3)$, i.e. the  
\underline{group of rotations} around the origin of ${\bf R}^3$, 
$$
SO(3) = \{ 3\times 3 \; \mbox{matrices} \ A \ 
\mbox{with real
coefficients} \ | \; 
A^t A = {\bf 1}_3 , \, {\rm det}\, A =1 \}
\ \ , 
$$
the group operation being the multiplication of matrices. 
This group contains in particular
the rotation $R_3 (\varphi )$ 
by an angle $\varphi \in {\bf R}$ around the 
$z$-axis, 
$$
\left[
\begin{array}{c}
x^{\prime} \\ y^{\prime} \\ z^{\prime} 
\end{array} 
\right] 
= \left[ 
\begin{array}{ccc} 
\cos  \varphi & - \sin  \varphi & 0 \\
\sin  \varphi & \cos  \varphi & 0 \\ 
0 & 0 & 1 
\end{array}
\right] \left[ 
\begin{array}{c}
x \\ y \\ z 
\end{array} 
\right] 
\equiv R_3 (\varphi ) \left[ 
\begin{array}{c}
x \\ y \\ z 
\end{array} 
\right] 
$$
as well as the analogous  
rotations around the $x$- and $y$-axis : 
$$
R_1 (\varphi) = 
\left[ 
\begin{array}{ccc} 
1 & 0 & 0 \\
0 & \cos  \varphi & - \sin  \varphi  \\
0& \sin  \varphi & \cos  \varphi \\ 
\end{array}
\right] 
\quad , \quad 
R_2 (\varphi )=  \left[ 
\begin{array}{ccc} 
\cos \, \varphi & 0 &  \sin \, \varphi \\
0 & 1 & 0 \\
- \sin \, \varphi & 0 & \cos \, \varphi \\ 
\end{array}
\right] 
\ \ . 
$$
A rotation by an angle $\varphi$ around an arbitrary direction in space 
 can be characterized by 
a vector $(\alpha , \beta , \gamma ) \in {\bf R}^3$
pointing in this direction and having a length equal to $\varphi$. 
 Thus, any element of $SO(3)$ is 
labeled by three real parameters :  this Lie group is 
three-dimensional and it is non-abelian, because rotations 
do not commute in general.   
Another example of a {\em matrix Lie group}  is given by the 
special unitary group $SU(n)$ defined in section 3.3.   
These matrix groups and ${\cal T} _n$ all represent 
examples of {\em Lie transformation groups}, i.e. groups
whose elements act on a certain space.   
    
To every Lie group one can associate its {\em Lie algebra}. 
If the former consists of  transformations on a certain
space, the latter represents the set of {\em 
infinitesimal transformations} on this space. 
For concreteness, we
consider the rotation $R_3 (\varphi )$ 
by an angle $\varphi$ 
which is close to $0$ (i.e. $R_3(\varphi )$ is close to the 
identity rotation)
and expand it in a Taylor
series around $\varphi =0$ : 
\begin{eqnarray}
R_3 (\varphi ) &=& R_3 ( 0) + \left. \ds{dR_3 \over d\varphi}
\right \vert _{\varphi =0} \ \varphi \, + \, ...
\\
& =&  
{\bf 1}_3 + r_3 \; \varphi + ...
\ \ . 
\nonumber
\end{eqnarray}
By this way of reasoning, 
one finds that the rotations $R_1 (\varphi ), R_2 (\varphi ),
 R_3 (\varphi )$ admit  as `infinitesimal generators' the
 antisymmetric matrices \eq 
r_1 =  
\left[ 
\begin{array}{ccc}
0 & 0 & 0 \\
0 & 0 & -1 \\
0 & 1 & 0 
\end{array} \right]
\quad , \quad r_2 = 
\left[ 
\begin{array}{ccc}
0 & 0 & 1 \\
0 & 0 & 0 \\
-1 & 0 & 0 
\end{array} \right]
\quad , \quad 
r_3 = \left[ 
\begin{array}{ccc}
0 & -1 & 0 \\
1 & 0 & 0 \\
0 & 0 & 0 
\end{array} \right]
\ \ . 
\eqn{gen}
In terms of the \underline{Lie commutator} (of matrices)
\eq
\fbox{\mbox{$
\ {[ A , B ]} \equiv AB - BA 
\ $}}
\ \ , 
\eqn{lie}
we have 
\begin{equation}
\left.
\begin{array}{l}
{[ r_1 , r_2 ]} = r_3 \\
{[ r_3 , r_1 ]} = r_2 \\
{[ r_2 , r_3 ]} = r_1 
\end{array}
\right\}
\ \Longleftrightarrow
\ \fbox{\mbox{$
\ [ r_j , r_k ] = \sum_{l=1}^3 \varepsilon _{jkl} \, r_l
\ $}}
\ \ {\rm for} \ \; j,k \in \{ 1,2,3\} 
\ \ .
\end{equation}
Here, the so-called \underline{structure constants} 
$\varepsilon_{jkl}$ are the elements of the totally
antisymmetric tensor of rank three normalized by 
$\varepsilon_{123} =1$ (i.e.
$1= \varepsilon_{123}=  \varepsilon_{312}= 
\varepsilon_{231}= -\varepsilon_{213}=
-\varepsilon_{132} = -\varepsilon_{321}$ and $\varepsilon_{jkl}=0$
otherwise). 

By definition, the \underline{Lie algebra} $so(3)$ 
\underline{associated to the Lie group} $SO(3)$ 
(which is $3$-dimensional) 
is the real, $3$-dimensional  vector space consisting of all linear combinations 
of the matrices 
$r_1, r_2$ and $r_3$ together with the Lie bracket operation   
defined by \equ{lie}. 
Explicitly, we have 
$$
so(3) =
 \{ 3\times 3 \; \mbox{matrices} \ A \ 
\mbox{with real  
coefficients} \ | \; 
A^t  = -A  \}
\ \ . 
$$

Note that two Lie groups which ``look the same in the 
neighborhood of the identity" admit the 
same Lie algebra since the latter is only defined 
by considering such a neighborhood. E.g. 
the elements of ${\bf R} - \{ 0 \}$ and ${\bf R}_+$
which are close to the identity $1$  have the 
form  ${\rm e}^a = 1 + 1 \cdot a + ...$
with $a\in {\bf R}$ and therefore both Lie groups 
admit ${\bf R}$ as their Lie algebra. 

Lie algebras may also be introduced without any reference 
to Lie groups : 
a (real) \underline{abstract Lie algebra} 
is a (real) vector space ${\cal G}$ together with an operation 
$[\, , \, ] : {\cal G} \times {\cal G} \to {\cal G}$ which is (${\bf R}$-) bilinear, 
antisymmetric and satisfies the Jacobi identity (i.e. 
$ 0 = [A,[B,C]] \, +$ cyclic permutations of $A,B,C$ ).   
E.g. the vector space $\rr^3$ together 
with the cross product as 
commutator, 
$[ \vec a , \vec b \, ] = \vec a \times \vec b $, represents an abstract Lie algebra.  
Note that the vectors 
$\vec e _1 = (1,0,0), \vec e _2 = (0,1,0), \vec e _3 = (0,0,1)$ which define 
the canonical basis of ${\bf R}^3$ satisfy 
$\vec e _j \times \vec e _k = \sum_{l=1}^3 \varepsilon_{jkl} \, \vec
e _l$ for $j,k \in \{ 1,2,3\}$,  i.e. the same relation as the
matrices $r_j \in so(3)$.

While we passed from the Lie group $SO(3)$ to the associated     
Lie algebra $so(3)$ by {\em differentiation}, 
we can go the other 
way round by {\em exponentiation} : in fact, one can verify 
that 
$R_3 (\varphi ) = {\rm exp} \, (\varphi r_3)
= {\bf 1}_3 + \varphi r_3 + ...$ where 
the exponential of matrices is defined by the Taylor
series expansion
(${\rm exp} \, A \equiv \sum_{n=0}^{\infty} {1 \over n!} A^n$).   

To summarize, 
the sets of finite and infinitesimal
transformations are related as follows :  
$$ 
\fbox{\mbox{$ 
\ \begin{array}{l c c c r}
                  & &  {\rm differentiate}      & &  \\
{\rm Lie \ group} & &  \rightleftharpoons & & {\rm Lie\ algebra} \\
                  & &  {\rm exponentiate}       & &   
\end{array} 
\ $}}
$$ 

Note that, in general, 
the exponentiation of Lie algebra elements
does not allow to recover all 
Lie group elements.
E.g.
exponentiation of the Lie algebra ${\bf R}$ 
associated to the Lie group $(\rr - \{ 0 \}, \cdot )$ yields 
the Lie group 
$\{ {\rm e}^a \, | \, a \in {\bf R} \} = {\bf R}_+$, i.e. 
the component of ${\bf R} - \{ 0 \}$ 
which is connected to the identity $1$.

Before proceeding further, we recall that a \underline{homomorphism} 
between two sets which are supplemented 
with additional structures is a {\em structure
 preserving map}. Thus, a {\em homomorphism of groups} is a map 
between two groups which preserves 
the group multiplication. E.g. the groups $({\bf R}^n , +)$ 
and  $({\cal T}_n , \circ)$ 
can be related by the map  
\begin{eqnarray}
f \ : \ \  {\bf R}^n  & & \! \longrightarrow \ \;  {\cal T}_n  
\nonumber
\\
a&  & \! \longmapsto \ \; T_a
\ \ ,
\nonumber
\end{eqnarray}
which is a homomorphism, because 
$f(a+b) = T_{a+b} = T_a \circ T_b = f(a) \circ f(b)$. 
The groups $({\bf R}, +)$ and $({\bf R}_+ , \cdot)$ 
 may be related by the homomorphism $a \mapsto {\rm e}^a$. 

Analogously, a {\em homomorphism of Lie algebras} 
is a map which preserves the Lie algebra
 structure, i.e. a linear map preserving the Lie bracket. For
 instance,  the Lie algebra ${\bf R}^3$ (with the 
cross product as commutator)
and the matrix Lie algebra $so(3)$ can be  related 
by the following mapping  
of the basis vectors,   
\begin{eqnarray}
\nonumber 
f \ : \ \  {\bf R}^3  & & \! \longrightarrow \ \;  so(3) 
\\
\vec e _j&  & \! \longmapsto \ \; r_j 
\ \ , 
\nonumber 
\end{eqnarray}
which we  
extend in a  linear way to all linear combinations 
of vectors  : this map is a homomorphism 
since it is linear (by definition) and 
$$
[ f( \vec e_j ), f (\vec e _k) ] = [ r_j, r_k ] = \sum_{l=1}^3 
\varepsilon_{jkl} \, r_l =   \sum_{l=1}^3 
\varepsilon_{jkl} \, f(\vec e_l ) = f( \sum_{l=1}^3 
\varepsilon_{jkl} \, \vec e_l ) = f (\vec e _j \times \vec e _k) 
,
$$
which implies $[ f( \vec a ), f (\vec b) ] = f  ( \vec a \times
\vec b )$ for all $\vec a, \vec b \in  {\bf R}^3$ by linearity. 

Actually, all of these illustrations represent 
examples of \underline{isomorphisms},
i.e. bijective maps such that the map and its inverse are homomorphisms.
Two groups, Lie algebras,... related by an isomorphism have 
exactly the 
same structure and thus {\em may be identified with each other}
 for many purposes. 

\newpage 
  
\subsection{About representations}

\vspace{-12mm}

\setcounter{subsubsection}{0}
\subsubsection{Representation of a (Lie) group}

\vspace{-2mm}

Suppose you want to let the elements of  the group $SU(2)$, i.e.  
complex $2\times 2$ matrices  
act linearly on vectors of $\rr ^3$  
in such a way that the group operation is preserved. 
More generally, you may want to define the action of a group $G$ 
on a real vector space $V$ of finite dimension $n$. 
The device which allows to realize  this idea is provided by
the next definition. 

Let $GL(V)$ denote the group of all invertible linear maps from $V$ 
to itself, the group operation being the composition of maps. 
This group is isomorphic to the group $GL(n, {\bf R})$ 
of all invertible $n\times n$ 
matrices with real coefficients, the group law being the matrix multiplication.   

\begin{defin}
A \underline{representation $D$ of the 
group $G$ 
on the vector space $V$ }  is a homomorphism  
from the group $G$ to the group  $GL(V)$, i.e.  it is a map 
\begin{eqnarray}
D \ : \ \  G & & \! \longrightarrow \ \; GL(V) 
\\
g &  & \! \longmapsto \ \; D(g) 
\nonumber 
\end{eqnarray}
 such that the linear operators $D(g)$ satisfy  
\eq
\fbox{\mbox{$
\ D(g_1 \circ g_2 ) = D(g_1) \circ D(g_2)
\ $}}
\quad for \ \, all \  g_1, g_2 \in G
\ \ . 
\eqn{repgr} 
The \underline{dimension of the representation} is the  
dimension of the representation space, i.e. 
${\rm dim} \, D = {\rm dim} \, V$.
\end{defin}
We now come back to the question 
concerning the action of $SU(2)$ on $\rr^3$ : 
rather than providing a complete answer (which may be found 
in many textbooks on quantum mechanics or quantum field 
theory), we  give an illustration for a particular element of 
$SU(2)$ :
\begin{eqnarray}
D \ : \ SU(2) & & \longrightarrow \ \ SO(3) \subset GL( 3, {\bf R}) 
\nonumber
\\
A_{\varphi} \equiv \left[ 
\begin{array}{cc}
{\rm e}^{{\rm i} \varphi /2} & 0 \\
0 & {\rm e}^{-{\rm i} \varphi /2}
\end{array} \right] 
& & \longmapsto \ \ 
D(A_{\varphi}) =  \left[ 
\begin{array}{ccc}
\cos \varphi & -\sin \varphi & 0 \\
\sin \varphi & \cos \varphi & 0 \\
0 & 0 & 1 
\end{array} \right] 
\ \ . 
\nonumber 
\end{eqnarray}
Here, $\varphi$ is a real parameter and it is straightforward 
to verify that the group product is preserved by the mapping $D$, i.e.
$
D (A_{\varphi_1} A_{\varphi_2} ) =    
D (A_{\varphi_1}) D( A_{\varphi_2} )$.

The previous definition can be generalized to the cases where $V$ 
is a complex vector space or where its dimension is not finite. 
In the latter case, one encounters the technical complication 
that the linear operators acting on $V$ can, in general, only be
defined on a  subspace of $V$ (e.g. the action of differential
ope\-rators on  function spaces can only be defined on sufficiently
smooth functions). However, in physics one usually deals with vector
spaces that are supplemented with a Hilbert space structure, e.g. 
the (complex, infinite-dimensional) Hilbert  space of
square-integrable functions on ${\bf R}$,  
 \eq
{\cal H} \equiv {\cal L}^2 ({\bf R}) 
= \{ \psi : {\bf R} \to {\bf C} \ | \; 
\int_{-\infty}^{+\infty} |\psi (x)|^2 dx < \infty \} 
\eqn{qm}
describing the quantum mechanical states of a particle 
moving on the real axis. In this case, one generally considers 
{\em unitary representations}, i.e. the group representative 
$D(g)$ of each element $g \in G$ is a unitary operator on 
the Hilbert space ${\cal H}$ (and thereby defined on all vectors
of ${\cal H}$). 

Let us illustrate this notion by an example involving the space
\equ{qm}.  Suppose you want to 
translate the wave function 
$\psi \in {\cal H}$ of a particle 
moving on the real axis by an amount $a \in \rr$. 
In other words, you want to consider the action of the 
translation $T_a \in {\cal T}_1$ on the vector $\psi$ :    
the group element  $T_a$ acts by means of the Hilbert space operator
\begin{eqnarray}
D(T_a) \ : \ \  {\cal H} & &
\! \stackrel{{\rm linear}}{\longrightarrow}
 \ \; {\cal H}
\nonumber 
\\
\psi &  & \! \longmapsto \ \; D(T_a) \psi \equiv \psi_a 
\nonumber  
\end{eqnarray}
defined by 
\eq
\psi_a (x) := \psi (x-a) 
\ \ . 
\eqn{transw}

\bigskip 

\begin{figure}[h!]
\centerline{\includegraphics*[height=3cm,angle=0]{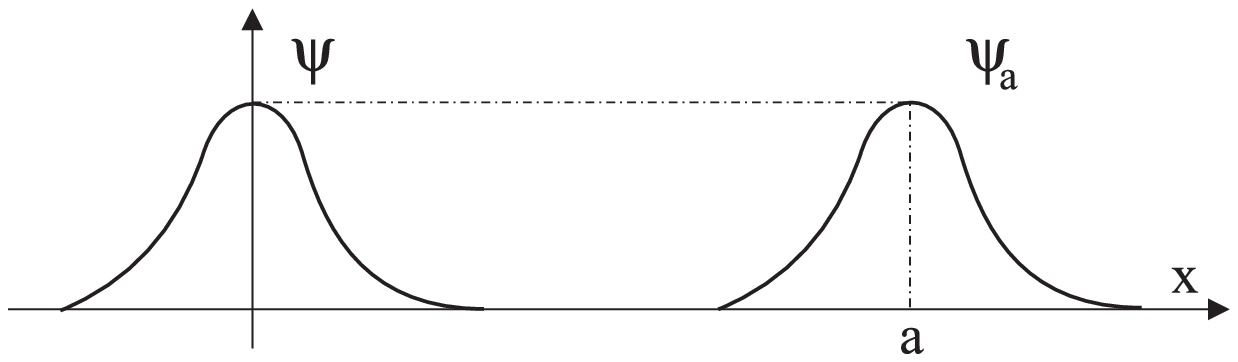}}
\end{figure}

\bigskip

\noindent The operator $D(T_a)$ is unitary and 
it defines a representation since 
$$
D(T_a \circ T_b ) \, \psi =
D(T_{a+b} ) \, \psi = \psi _{a+b} = D(T_a) \, \psi_b 
= D(T_a)  D(T_b) \, \psi 
\ \ {\rm for} \ {\rm all} \ \ \psi \in {\cal H}
\,  , 
$$
henceforth 
$D(T_a \circ T_b ) = D(T_a)  D(T_b)$.  
The dimension of this unitary representation   
of the translation group ${\cal T}_1$ is infinite.

\subsubsection{Representation of a Lie algebra}

\vspace{-2mm}

As in the previous section, we consider a real vector space $V$ of
finite dimension. 
The set  $gl(V)$ consisting of {\em all} linear operators from $V$ to
itself, 
equipped with the usual linear  operations (addition of
operators and multiplication  of operators 
by real numbers) and with the commutator
$[A, B] =A  \circ B - B  \circ A$ is a real Lie algebra : it is the 
Lie algebra associated to the Lie group  $GL(V)$.

\begin{defin} 
A \underline{representation $d$ of the 
(real) Lie algebra ${\cal G}$ 
on the vector space $V$} is a homomorphism from the 
Lie algebra ${\cal G}$ to  the Lie algebra $gl(V)$, i.e. it is a map 
\begin{eqnarray}
d \ : \ \  {\cal G} & & \! \longrightarrow \ \; gl(V) 
\\
t &  & \! \longmapsto \ \; d(t) 
\nonumber 
\end{eqnarray}
such that the linear operators 
$d(t)$ satisfy 
 \eq
\fbox{\mbox{$
\ d(c_1 t_1 + c_2 t_2 ) = c_1 d(t_1) + c_2 d(t_2) 
\ $}} 
\quad and \quad 
\fbox{\mbox{$
\ d( [t_1, t_2] ) = [d(t_1) , d(t_2) ]  
\ $}}
\eqn{repal} 
for all $t_1, t_2 \in {\cal G}$ and  $c_1,  c_2 \in {\bf R}$. 
Furthermore, one defines ${\rm dim} \,  d =
{\rm dim} \, V$ \footnote{\small 
If the representation space is an infinite-dimensional  Hilbert
space ${\cal H}$, one assumes in addition that the operators  
$d(t)$ for all $t \in {\cal G}$ have a
common invariant domain ${\cal D}$ which is dense in  ${\cal
H}$ \cite{barut}.}. 
\end{defin}

As an illustration, we consider the 
operators $L_1, L_2,  L_3$ of angular momentum
which act on the Hilbert space 
${\cal H} = {\cal L}^2 ({\bf R}^3)$ of quantum mechanics : they
admit  a common  invariant domain ${\cal D}$ 
which is dense in  ${\cal H}$ and on which they 
satisfy the algebra  
$$ [ L_j , L_k ] = {\rm i}
\hbar \, \sum_{l=1}^3 \varepsilon_{jkl} \, L_l  \qquad \quad {\rm
for} \ \; j,k \in \{ 1,2,3\}  
\ \ . 
$$ 
Therefore,  
the mapping 
\begin{eqnarray}
d\ : \ & so(3) & \longrightarrow \ gl({\cal D})  
\nonumber 
\\
& r_k& \longmapsto \ d(r_k) \equiv {1 \over {\rm i} \hbar} \, L_k 
\end{eqnarray}
(extended by linearity to all elements of the vector space 
$so(3)$) is a representation of the Lie algebra of infinitesimal 
rotations on the Hilbert space ${\cal H}$. 
(We leave the verification  of the properties \equ{repal}
as an exercise.)

The image $D(G)$ of a Lie group $G$ under a representation 
$D$ is again a Lie group 
and similarly $d({\cal G})$ is a Lie algebra.    
The passage from the set $D(G)$ of finite transformations
to the set $d({\cal G})$ of infinitesimal transformations  
can be performed 
in the same way as the one between $G$ and ${\cal G}$, i.e. 
by differentiation and exponentiation, respectively.


\end{document}